\documentclass[12pt]{amsart}

\newcommand{\titel}{A Practical Algorithm for Reconstructing 
Level-1 Phylogenetic Networks}
\usepackage{ifthen}

\usepackage{epic}
\usepackage[leqno]{amsmath}

\usepackage{latexsym} 
\usepackage{amsmath} 
\usepackage{epsfig}
\usepackage{a4wide}
\usepackage{amssymb} 
\usepackage{enumerate}
\usepackage{url}

\usepackage[mathscr]{euscript}

\newtheorem{theorem}{Theorem}

\usepackage{algorithm}
\usepackage[noend]{algorithmic}
\usepackage{epsfig}

\newif\ifcomment\commentfalse
\def\commentON{\commenttrue}

\long\outer\def\bc#1\ec{{\ifcomment \sloppy  $[${\bf Steven says}]
{{#1}} \textbf{[end]} \fi }}

\long\outer\def\br#1\er{{\ifcomment \sloppy  $[${\bf Steven suggests to remove}]
{{#1}} \textbf{[end]} \fi }}

\long\outer\def\bo#1\eo{{\ifcomment \sloppy  $[${\bf instead of}]
{\textit{#1}} \textbf{[end]}  \fi }}

\long\outer\def\bn#1\en{{\ifcomment \sloppy  $[${\bf New suggestion}]
{{#1}} \textbf{[end]} \fi }}

\long\outer\def\BC#1\EC{{\ifcomment \sloppy \par \#  \dotfill
{\textsc{#1}} \dotfill \# \par \fi }}

\commentON

\begin{document}

\title{\titel}

\author{Katharina T. Huber, Leo van Iersel, Steven Kelk and Rados{\l}aw Suchecki}

\thanks{Huber and Sucheki are affiliated
with the School of Computing Sciences, University of East Anglia, Norwich, NR4 7TJ, United Kingdom, email:
\texttt{Katharina.Huber@cmp.uea.ac.uk, R.Suchecki@uea.ac.uk}. Van Iersel is affiliated with University of Canterbury, Department of Mathematics and 
Statistics,
Private Bag 4800, Christchurch, New Zealand, email: \texttt{l.j.j.v.iersel@gmail.com}. Kelk is affiliated with the Centrum voor Wiskunde en Informatica (CWI),
P.O. Box 94079, 1090 GB Amsterdam, The Netherlands, email: \texttt{s.m.kelk@cwi.nl}.}

\maketitle

\begin{abstract}
Recently much attention has been devoted to the construction of phylogenetic networks which generalize 
phylogenetic trees in order to accommodate complex evolutionary processes. Here we present an efficient, practical 
algorithm for reconstructing ${\mbox{level-}1}$ phylogenetic networks - a type of network slightly more general
than a phylogenetic tree - from triplets. Our algorithm has been made publicly available as the program \textsc{Lev1athan}. It
combines ideas from several known theoretical algorithms for phylogenetic tree and network reconstruction with two novel 
subroutines. Namely, an exponential-time exact and a greedy algorithm both of which are of independent theoretical 
interest. Most importantly, \textsc{Lev1athan} runs in polynomial time and always
constructs a level-1 network. If the data is consistent with a phylogenetic tree, then the algorithm constructs such a tree.
Moreover, if the input triplet set is dense and, in addition, is fully consistent with some 
${\mbox{level-}1}$ network, it will find such a network. The potential of \textsc{Lev1athan} is explored by means of an extensive
simulation study and a biological data set. One of our conclusions is that \textsc{Lev1athan} is able to construct networks
consistent with a high percentage of input triplets, even when these
input triplets are affected by a low to moderate level of noise.
\end{abstract}


\section{Introduction}

\begin{figure}[t]
\centering
\includegraphics[scale=.5]{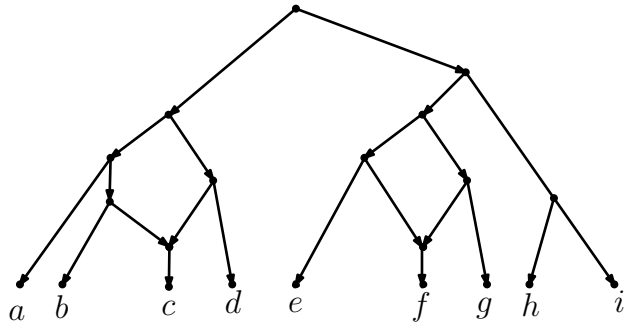}
\caption{A phylogenetic network with leaf set $\{a,\ldots,i\}$ 
in the form of a level-1 network.}
\label{fig:phylogenetic network}
\end{figure}

Phylogenetic networks such as the one depicted in 
Figure~\ref{fig:phylogenetic network} provide a natural and powerful
extension of the concept of a phylogenetic tree 
(see Section~\ref{subsec:def} for precise 
definitions of these two concepts as well as the other concepts used in 
this paper) to accommodate complex evolutionary processes such as 
hybridization, recombination or horizontal gene transfer. 
Consequently, their attractiveness to evolutionary biology  as
a model for representing the evolutionary past of a set of
taxa (e.g. species represented by gene or genetic marker sequences) whose
evolution might have been driven by such processes
has grown over the years. This in turn has generated much interest in these structures from mathematicians
and computer scientists working in phylogenetics \cite{Gambette2009,Huson2007, HusonRupp2009,Semple2007}.
The desire of biologists to use ever-longer sequences, combined with
the computational complexities involved in dealing with such sequences,
has meant that much research in mathematical methodology 
and algorithm development has focused on developing indirect
methods for phylogenetic reconstruction. This has spawned the thriving area of supertree construction
\cite{BinindaEmonds2004} which aims to reconstruct a phylogenetic tree by puzzling it together
from smaller trees. Inspired by this - and guided by 
the fundamental observations
that (i) a phylogenetic tree is uniquely described by the
set of triplets (i.e. rooted binary phylogenetic trees on three leaves) it induces (see e.g. \cite{SempleSteel2003}) and (ii) that
a phylogenetic network is a generalization of a phylogenetic tree,
two main triplet-based approaches for phylogenetic network reconstruction
immediately suggest themselves. One is to essentially first employ a
method such as TreePuzzle (see e.g. \cite{SchmidtEtAl2002}) to generate a set of
 quartets\footnote{The analogue of a triplet within the
area of unrooted phylogenetic tree reconstruction.} 
from a sequence alignment, to then derive a set $T$ of triplets from
that set and then to use the set $T$ to reconstruct a phylogenetic 
network. The other is to essentially first reconstruct phylogenetic trees
on subsections of the given alignment (that might reflect different evolutionary scenarios)
and then to reconstruct a phylogenetic network from the set of triplets induced by those trees.

For any set $T$ of triplets, a phylogenetic network that, in a well-defined
sense, is {\em consistent} with 
all triplets in $T$ can easily be constructed if the complexity of the network is
essentially unbounded~\cite{JanssonSung2006, reflections}. 
Such networks are however only of marginal biological
relevance. Researchers have therefore turned their attention to studying restricted 
classes of phylogenetic networks. One such class is that of a 
\emph{${\mbox{level-}1}$ network}
or a \emph{galled tree} i.e. a phylogenetic network in which essentially
any two cycles are vertex disjoint (see Figure~\ref{fig:phylogenetic network}
for an example). Such networks are of practical interest because (i) 
they are relatively treelike, and (ii) their simple structure suggests the possible existence of fast algorithms to construct them. Indeed, Jansson, Sung and Nguyen showed that it can be 
decided in polynomial time whether there exists a ${\mbox{level-}1}$ network with leaf set $X$ consistent with all input triplets~\cite{JanssonEtAl2006,JanssonSung2006}, if the input 
triplet set is \emph{dense}, i.e. if a triplet is given for each combination of three taxa in $X$. Their algorithm will also construct such a network, if it exists. However, a 
${\mbox{level-}1}$ network consistent with all input triplets might not exist for several reasons. Firstly, the real evolutionary history might be too complex to be described by a 
${\mbox{level-}1}$ network. Secondly, some of the input triplets might be incorrect (which is likely to be the case in practice).

One response to this problem has been to increase the complexity
of the networks that can be modelled. For example, it has 
been shown that, for each fixed non-negative integer $k$, the problem
of constructing a level-$k$ network consistent with a dense set of 
input triplets is polynomial-time solvable
\cite{tohabib2009,lev2TCBB}. The higher the level, the higher the 
complexity of evolutionary scenarios that can be represented.
However, the running time grows exponentially with $k$ and initial experiments 
with the related heuristic 
\textsc{Simplistic} \cite{simplicityAlgorithmica,SIM08} show that, since these algorithms insist on full consistency with the input triplets, only a small amount of noise is required in 
the input data to artificially inflate the level of the produced network. This causes an undesirable increase in both running time and network complexity.

A second strategy is to place a ceiling on the complexity of the networks that can be constructed and to no longer
demand full triplet consistency. Implemented in the program 
\textsc{Lev1athan}~\cite{lev1athandownload},
this paper adopts this second strategy and presents the first heuristic algorithm for 
constructing ${\mbox{level-}1}$ networks from triplets. Given any set of triplets, our heuristic always
constructs a ${\mbox{level-}1}$ network $\mathcal N$ in polynomial time, which is in practice a great advantage
in light of the algorithmic results mentioned above. Moreover, it attempts to construct $\mathcal N$ such that 
it is consistent with as many of the input triplets as possible. 
If a weight is given for each triplet reflecting for example some kind of confidence level one might have in that
triplet, then our heuristic aims 
to maximize the total weight of the input triplets consistent with $\mathcal N$. Optimizing such functions 
is NP-hard~\cite{reflections,Wu2004}, even for the case of determining a tree consistent with
the maximum 
number of
triplets from an unweighted, dense triplet set.
Having said that, an 
exponential-time exact algorithm was proposed in~\cite{reflections} that always finds an optimal solution, but this
algorithm is only practical for small numbers of taxa. In addition, polynomial-time approximation algorithms have been
formulated for level-1 network reconstruction. For example it was first shown in \cite{JanssonEtAl2006} how to construct in polynomial time such a network consistent with 5/12 $\approx 0.42$ of the input triplets. This was subsequently improved to 
$0.488\ldots$ in~\cite{ByrkaEtAl2008}, which is worst-case optimal. Both algorithms are mathematically interesting, but have the drawback that they 
produce networks with a highly rigid topology which are biologically 
unrealistic. 

\textsc{Lev1athan}, which we outline in Section \ref{sec:method},
combines elements from the above mentioned approaches into an algorithm with
a strong recursive element. In addition to its polynomial running time,
\textsc{Lev1athan} enjoys the following desirable properties. If a set of input
triplets is consistent with a tree, or if at any stage in a recursion such a triplet set $T$ occurs, a phylogenetic tree will be 
generated from $T$. Similarly, whenever the triplet set is dense and fully consistent with some ${\mbox{level-}1}$ network, such a 
network will be produced. Both outcomes are a direct consequence of a partitioning strategy that we describe in 
Section~\ref{sec:partition}.
If a network is produced that (when ignoring directions on the arcs) contains cycles of 
moderate size, then, due to a novel exponential-time exact algorithm
which we describe in Section \ref{sec:simple}, 
the topology of each of these cycles is locally optimal. This algorithm is complemented by a novel greedy algorithm
to construct larger cycles
which we also describe in that section. In addition, the output network is guaranteed to be consistent with at least 5/12 of the
input triplets, if one uses the \textsc{Lev1athan} option ``blocks 3'' which has its origin in the above mentioned partitioning strategy.
In Section \ref{sec:experiments} we test \textsc{Lev1athan} on synthetic and real biological data; to facilitate this we
also develop a novel ${\mbox{level-}1}$ network generation method and 
discuss several measures for network comparison. Taken as a whole the results of the experiments are promising, suggesting that
\textsc{Lev1athan} will be genuinely useful in a real-world context, but do also highlight some limitations of triplet-based approaches and level-1 networks as a model of
evolution.

We conclude this section with remarking that although
\textsc{Lev1athan} can be applied to both weighted and unweighted triplet 
sets for clarity we will restrict our exposition to unweighted triplet sets. 
\vspace{-0.3cm}
\section{Basic Concepts}
\label{subsec:def}
We start with some concepts from graph theory concerning directed
acyclic graphs.
Suppose $G=(V,A)$ is a directed acyclic graph, or DAG for short. Then $G$ is called  
\emph{connected} (also called
``weakly connected'') if, when ignoring the directions on the arcs, there is a path between any two vertices of $G$. Moreover, $G$ is
called \emph{biconnected} if $G$ contains no vertex whose removal disconnects $G$. A biconnected subgraph~$H$ of a graph~$G$ is said to be a \emph{biconnected component} if there is no biconnected subgraph~$H'\neq H$ of $G$ that contains~$H$. A biconnected component is said to be \emph{nontrivial} if it contains at least three vertices.
A vertex $v$ of $G$ is called an {\em ancestor} of a vertex $v'$ if there is a directed path from $v$ to $v'$. In this case, $v'$ is also called a {\em 
descendant} of $v$. Now suppose that $v_1$ and $v_2$ are two distinct vertices of $G$. Then a vertex $v\in V$ is a {\em lowest common ancestor} of $v_1$ and $v_2$ if $v$ is an ancestor of both $v_1$ and $v_2$ and no descendant of $v$ is also an ancestor of $v_1$ and $v_2$.

A \emph{(phylogenetic) network} $\mathcal{N}$ on some finite set $X$ of taxa is a DAG with a single root (a vertex with indegree~0), whose leaves (vertices 
with outdegree~0) 
are bijectively labelled by the elements of $X$. Following common practice, we identify each leaf with its label. Thus, the leaf set
of $\mathcal{N}$, denoted $L(\mathcal{N})$, is $X$. Vertices with indegree~1 are called {\em split vertices} and vertices with indegree at least two are 
called {\em 
reticulations}.
A network $\mathcal{N}$ is called {\em semi-binary} if each reticulation has indegree 2 and \emph{binary} if in addition each reticulation has outdegree 1 and each
split vertex outdegree 2. We will restrict 
our exposition to binary networks, but remark that \textsc{Lev1athan} uses post-processing to simplify a constructed network by collapsing arcs while making sure that triplet consistency is preserved. This can create vertices with outdegree greater than two.

\begin{figure}[H]
\centering
\includegraphics[scale=.6]{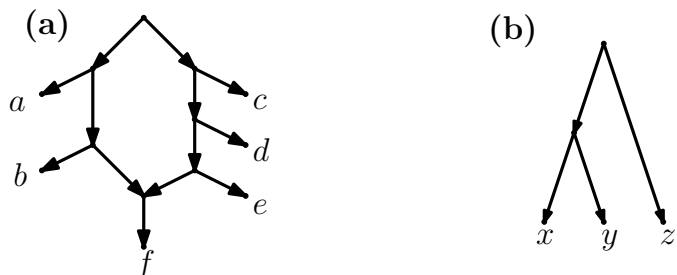}
\caption{ \label{fig:singletriplet}(a) Example of a simple level-1 network 
on $X=\{a,\ldots, f\}$ and (b)
 the triplet $xy|z$.}
\end{figure}

A semi-binary network $\mathcal{N}$ is said to be a \mbox{\emph{level-}$k$ \emph{network}}, if each biconnected component contains at most~$k$ reticulations. Clearly, if 
$k=0$ then this implies that $\mathcal{N}$ is a phylogenetic tree in the usual sense (see e.g.~\cite{SempleSteel2003}). It is not difficult to verify that for $k=1$, a nontrivial biconnected component of a level-$k$ network always has the form of two directed paths that both start at some vertex $u$ and both end at some vertex
$v \neq u$ but which are otherwise vertex disjoint. Such a biconnected component is often called a \emph{reticulation cycle} or a
\emph{gall}. We define the {\em size} of a gall to be the number of vertices in it i.e. the number of outgoing arcs plus one. 
The network in Figure~\ref{fig:singletriplet}(a) contains a single gall of size 7, for example. Let $\mathcal{N}$ be a phylogenetic network. An arc $a$ of 
$\mathcal{N}$ 
is said to be a 
\emph{cut-arc} if removing $a$ disconnects $\mathcal{N}$. A cut-arc is called \emph{trivial} if its head is a leaf. A level-$k$ network is said to be a 
\emph{simple} level-$k$ 
network, if $\mathcal{N}$ contains no nontrivial cut-arcs and is not a level-$(k-1)$ network. Thus, a simple ${\mbox{level-}1}$ network can be 
thought of as a single gall with some leaves attached to it, as in Figure~\ref{fig:singletriplet}(a). Simple ${\mbox{level-}1}$ networks are the basic, recursive building 
blocks of ${\mbox{level-}1}$ networks, in the sense that each gall of a ${\mbox{level-}1}$ network essentially corresponds to a simple ${\mbox{level-}1}$ network.

A \emph{(rooted) triplet}\index{triplet} $xy|z$ is a phylogenetic tree on $\{x,y,z\}$ such that the lowest common
ancestor of $x$ and $y$ is a proper descendant of the lowest common ancestor of $x$ and $z$, see Figure~\ref{fig:singletriplet}(b). For convenience, we call a set of triplets a {\em triplet set}. A triplet $xy|z$ is said to be \emph{consistent} with a network $\mathcal{N}$
(interchangeably: $\mathcal{N}$ is consistent with $xy|z$) if $\mathcal{N}$ contains a subdivision of
$xy|z$, i.e. if $\mathcal{N}$ contains distinct vertices $u$ and $v$ and pairwise internally vertex-disjoint
paths $u \rightarrow x$, $u \rightarrow y$, $v \rightarrow u$ and $v\rightarrow z$. For example, the network in 
Figure~\ref{fig:singletriplet}(a)
is consistent with (among others) the triplets~$ab|c,af|c$, $ef|d$ and~$de|f$ but is not consistent with (among others) the triplets~$be|a,cd|e$ and~$fa|b$. We say that a triplet set $T$ 
is 
consistent with $\mathcal{N}$ if all triplets in $T$ are consistent with $\mathcal{N}$ and denote the set of all triplets on $L(\mathcal N)$ consistent with $\mathcal{N}$ by $T(\mathcal{N})$.
\vspace{-0.2cm}
We note that, whenever a network $\mathcal{N}$ contains a gall $C$ of size 3,
$C$ can be replaced by a single split node to obtain the 
network $\mathcal{N}'$ where $T(\mathcal{N}') = 
T(\mathcal{N})$. In such cases we prefer the latter construction to the gall because
it is a more parsimonious response to the input data. We henceforth sharpen
the above definition of a level-1 network to exclude galls of size 3.
(Galls of size 2 are already excluded because we do not allow multiple arcs).
\section{A brief outline of \textsc{Lev1athan}}
\label{sec:method}

In this section with present a rough outline of our 
4 phase program \textsc{Lev1athan} which is implemented in Java and 
freely available for download \cite{lev1athandownload}. It 
takes as input a set $T$ of triplets or, more generally,
a set $\mathcal T$ of phylogenetic trees 
(e.g. gene trees) given in Newick format \cite{newick}.
In the latter case the (weighted) union of the triplet 
sets induced by the trees in $\mathcal T$ is taken by \textsc{Lev1athan} as triplet set $T$. 
It outputs a (possibly post-processed) level-1 network 
in DOT format \cite{dot} and/or eNewick format \cite{eNewick}.
The goal of \textsc{Lev1athan} is to construct a level-1 network $\mathcal N$
that maximizes the number of triplets in $T$ that 
are consistent with it.
In an optional post-processing phase the generated
network can then be simplified by contracting all arcs $(u,v)$ of $\mathcal N$
where neither $u$ nor $v$ is a reticulation, $v$ is not a leaf, such that 
the contraction would not cause a triplet in $T$ that was consistent with 
$\mathcal N$ to become not consistent with it.
(Leaving such arcs uncontracted is tantamount to an
unfounded strengthening of our hypothesis about what the ``true'' evolutionary scenario 
looked like). To explain the 
program's four phases, we 
require some more definitions and notation. To this end, suppose $T$ 
is a set of triplets and define the {\em leaf set} $L(T)$ of
$T$ to be the set $\bigcup_{t\in T}L(t)$. For any subset
$L'\subseteq L(T)$, we denote by $T|L'$ the set 
of triplets in $t\in T$ such that $L(t)\subseteq L'$.

\begin{algorithm}[H]\caption{Basic outline of \textsc{Lev1athan}}\label{alg:leviathan}
\begin{algorithmic} [1]
\REQUIRE : Triplet set $T$.   
\ENSURE: Level-1 network $\mathcal{N}$ heuristically optimizing the number of triplets consistent with $T$.
\STATE \textbf{Partitioning the leaf set:} 
Find a partition~$\mathcal{L} = 
\{L_1,\ldots,L_q\}$ of $L(T)$. This will be
detailed in Section~\ref{sec:partition}. %
\STATE \textbf{Gall construction:} Construct a simple level-1 
network~$\mathcal{N}$ with 
leaf set $\mathcal{L}$. This will be detailed in 
Section~\ref{sec:simple}. %
\STATE \textbf{Recursion:} Recursively call \textsc{Lev1athan} to construct, for all
$1\leq i\leq q$, a level-1 network $\mathcal{N}_i$ from the triplet set
$T_i = T | L_i$.
\STATE \textbf{Merging:} Construct a network~$\mathcal{N}$ by 
combining~$\mathcal{N}$ with~$\mathcal{N}_1,\ldots,\mathcal{N}_q$ as follows.
For~$i=1,\ldots,q$, identify leaf~$L_i$ of~$\mathcal{\mathcal{N}}$ with 
the root of~$\mathcal{N}_i$. 
\end{algorithmic}
\end{algorithm}

\noindent Note that the algorithm does not backtrack 
in the sense that it never revises 
earlier made decisions. Also note that while the recursion and merging 
phases are relatively simple, the partition and gall 
construction phases are not. For these phases we designed new algorithms, 
which will be described in the following two sections.

\section{Partitioning the Leaf Set}
\label{sec:partition}

Based on their order of priority, we next describe the 
three steps that make up the partitioning strategy employed
by \textsc{Lev1athan} to find a suitable partitioning $\mathcal L$ of the
leaf set of a triplet set.
To explain this strategy in detail, assume for the rest of this section
that $T$ is a set of triplets and put $L:=L(T)$.

The first step of \textsc{Lev1athan}'s partitioning strategy is to determine 
whether an \emph{Aho move} is possible for $T$. This move is based on an
algorithm originally introduced by Aho et. al. in \cite{AhoEtAl1981}.
Following \cite{SempleSteel2003} where this algorithm is referred to as 
\textsc{Build} algorithm, this algorithm relies on the {\em clustering graph} 
$G_{[L',T]}$ induced by $T$ on any subset $L'\subseteq L$. For the 
convenience of the reader, we remark that for any subset $L'\subseteq L$
the vertex set of $G_{[L',T]}$
is $L'$ and any subset $\{a,b\}\in {L'\choose 2}$ forms an edge in $G_{[L',T]}$
if there exists some $c\in L'$ such that $ab|c\in T$.
\\

\noindent\textbf{Aho move:} Construct the clustering graph 
$G_{[L,T]}$. If  $G_{[L,T]}$
is not connected then use the vertex sets of the 
connected components of $G_{[L,T]}$ as the
 blocks of the partition $\mathcal{L}$. Otherwise, say that the Aho move
is not successful.\\

\noindent Note that the Aho move is essentially the embedding of the \textsc{Build} 
algorithm inside \textsc{Lev1athan}. Always starting
with such a move means that, if a set of triplets is consistent 
with a phylogenetic tree, then a phylogenetic tree will be output. More
generally it means that \textsc{Lev1athan} will potentially produce 
level-1 networks with treelike regions. Also note that if the
Aho move is successful then the \emph{Gall construction} phase 
of \textsc{Lev1athan} is not necessary: the components of the
clustering graph will simply correspond to subnetworks that the algorithms
``hangs'' from a single split vertex, just like the \textsc{Build} algorithm. (This can 
lead to split vertices with more than two children, which as already 
mentioned \textsc{Lev1athan} supports via post-processing).

\begin{figure}[H]
\centering
\includegraphics[scale=.5]{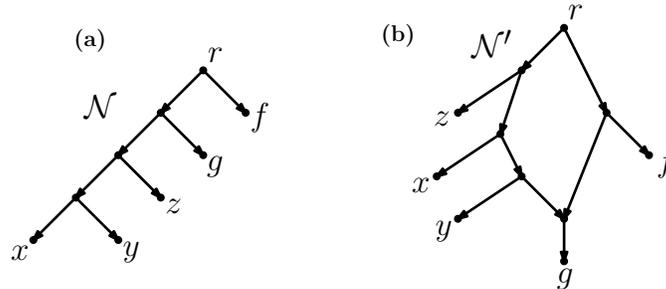} 
\caption{The triplets in $T(\mathcal{N})$ are also consistent with 
$\mathcal N'$
and $\mathcal N'$ is minimal with this property.}
\label{symdiff}
\end{figure}

Our motivation for prioritising
the Aho move is twofold. Firstly, it adheres to the 
parsimony principle: 
if the data can be explained equally well 
by both a phylogenetic tree and a network, choose the tree. As a non-trivial example,
note that the triplet set $T(\mathcal{N})$ induced by the  
phylogenetic tree
$\mathcal N$ on $X=\{x,y,z,g,f\}$
depicted in Figure~\ref{symdiff}(a) is also consistent with the
level-1 network $\mathcal N'$ on $X$ depicted in Figure~\ref{symdiff}(b) 
and that $\mathcal N'$ is in some sense minimal with this property, that is,
no arc of $\mathcal N'$ can be deleted such that the triplets in 
$T(\mathcal{N})$ remain consistent with the resulting phylogenetic tree. 
Secondly it exploits the not entirely trivial observation that, when attempting 
to construct a level-1 network consistent with the maximum number of triplets
in a given set, it can never be suboptimal to make an Aho move, when this is 
possible \cite{SnirRao2006}. 
%

If the Aho move is not successful, the next step is to try a \emph{JNS move}.
To explain this move we require a further concept which was originally 
introduced by Janson, Nguyen and Sung in \cite{JanssonEtAl2006}. 
Suppose $L'\subseteq L(T)$. Then $L'$ is called an {\em SN-set (of $T$)} 
if there exists no triplet $xy|z$ in $T$ such 
that $x, z \in L'$ and $y \not \in L'$.
A SN-set $S$ of $T$ is called \emph{maximal} if $S \neq L(T)$ and there does not exist
an SN-set $S' \neq L(T)$ such that $S \subset S'$.
Also, for some partition $\mathcal{L} = \{ L_1, ..., L_q \}$ of 
$L(T)$ into $q$ blocks $L_i$, $1\leq i\leq q$, we define the
\emph{induced} triplet set $T\nabla \mathcal{L}$  of $\mathcal L$ as the set 
of all triplets $L_i L_j |L_k$ on $\mathcal L$ for which there 
exists a triplet $xy|z \in T$ where $x \in L_i, y \in L_j$ and $z \in L_k$ and
$i$, $j$ and $k$ are all distinct.
Note that $T\nabla \mathcal{L}$ is dense if $T$ is dense.\\
\\
\noindent\textbf{JNS move:} If (a) the set $\mathcal{S}$ 
of maximal SN-sets of $T$ is a partition of $L(T)$, and (b)
$T \nabla \mathcal{S}$ is a dense set of triplets that is consistent with
some simple level-1 network $\mathcal{N}^{s}$,
then define $\mathcal{S}$ to be the sought after partition $\mathcal{L}$
(and use $\mathcal{N}^{s}$ as $\mathcal{N}$ in the Gall construction phase).
Otherwise and as in the case of an Aho move, say that the move is not
successful.\\

\noindent The JNS move is essentially the level-1 analogue of 
the above described
Aho move. The main difference is that although, using for example the
\textsc{Build} algorithm, it is always possible to decide in
polynomial time if a triplet set is consistent with a phylogenetic 
tree it is in general NP-hard
to determine whether there is a level-1 network that is consistent 
with the set (even if we restrict to simple level-1 networks). 
However, the situation changes and becomes
decidable in polynomial time if the triplet set
in question is dense  \cite{JanssonEtAl2006}. 
Density of the induced triplet set is therefore
a necessary requirement for a successful JNS move. Hence, requirement (b).
To motivate requirement (a), note that for dense triplet sets $T$
the set $\mathcal{S}$ of maximal SN-sets of $T$ always forms a partition
of $L(T)$. For general (i.e. non-dense) triplet sets this is not always, but
\emph{sometimes}, true. Requirement (a) is thus included to extend JNS moves to
non-dense triplet sets.

We conclude the discussion of the JNS move with remarking that this move 
enjoys the same optimality properties as an Aho move. More precisely, 
prioritising the JNS move guarantees that a level-1 network consistent with all 
the original triplets will be produced if such a network exists and the 
original triplet set was dense. (And, less obviously, that making a JNS move
can never lead to a suboptimal network, assuming subsequent recursive steps give
optimal networks). Again analogous to 
a successful Aho move, a successful JNS move allows the full generality of 
the Gall construction phase (see below) to be bypassed by utilizing the 
already computed simple level-1 network $\mathcal{N}^{s}$.

Due to noise in real biological data (or the inherent complexity of the underlying network) however, it is generally 
too much to hope for that one of the two moves described so far will
always be successful. We therefore have adapted a strategy 
from \cite{JanssonEtAl2006} to obtain a third move which
we call the {\em Heuristic move}. 
The heart of this move is formed by a score function 
$score(T,\mathcal{L'})$, $\mathcal {L'}$ a
partition of the leaf set of $T$, from 
\cite{JanssonEtAl2006} plus two operations (P1) and (P2)
that will allow us to manipulate a partition $\mathcal{L}_{old}$
with the aim of ensuring that the score of the new partition 
$\mathcal{L}_{new}$
is at least as good as the score of the given partition, i.e.
$score(T,\mathcal{L}_{new})\geq score(T,\mathcal{L}_{old})$ holds. To 
describe both the score function and these two operations, 
we require some more concepts. 
Suppose $\mathcal{L'} = \{L'_1, ..., L'_q\}$ is a partition of $L(T)$.
Then to $\mathcal{L'}$ we associate the following four subsets
\begin{align*}
T_{bad}&=T_{bad}(\mathcal{L'},T)
= \bigg \{ xy|z \in T \bigg | x, z \in L'_i, y \in L'_j 
\text{ where } i \neq j \bigg \},\\
T_{good}&=T_{good}(\mathcal{L'},T) 
= \bigg \{ xy|z \in T \bigg | x, y \in L'_i, z \in L'_j 
\text{ where } i \neq j \bigg \},\\
T_{local}&=T_{local}(\mathcal{L'},T)
= \bigg \{ xy|z \in T \bigg | x, \in L'_i, y \in L'_j, z \in L'_k
\text{ where } i \neq j\neq k\neq i \bigg \},\\
T_{defer}&=T_{defer}(\mathcal{L'},T) 
= \bigg \{ xy|z \in T \bigg | x,y,z \in L'_i 
\text{ for some } 1 \leq i \leq q \bigg \}.
\end{align*}
Note that $\{T_{bad},T_{good},T_{local},T_{defer}\}$ clearly
forms a partition of $T$. The aforementioned score 
function $score(T,\mathcal{L'})$ is then
defined as
$score(T,\mathcal{L'}) = 4|T_{defer}| + 7|T_{local}| + 12|T_{good}|$. 
Denoting as above a given partition by $\mathcal{L}_{old}$
and the generated partition by $\mathcal{L}_{new}$
then the two aforementioned operations (P1) and (P2)
are defined as follows.
\begin{enumerate}
\item[(P1)] If $A,B\in \mathcal{L}_{old}$ and $a\in A$ then
$\mathcal{L}_{new}=\{A-a,B\cup\{a\}\}\cup (\mathcal{L}_{old}-\{A,B\})$.
\item[(P2)] If $A\in \mathcal{L}_{old}$ with $|A|\geq 2$ and $a\in A$
then $\mathcal{L}_{new}=\{A-a,\{a\}\}\cup (\mathcal{L}_{old}-A)$.
\end{enumerate}
Armed with these definitions, we are now ready to state the Heuristic 
move.\\

\noindent\textbf{Heuristic move:} Starting with
$\mathcal{L}_{old} = \{ L \}$ we exhaustively search for an operation
(P1) or (P2) which, when applied to 
$\mathcal{L}_{old}$, will create a partition $\mathcal{L}_{new}$
with  $score(T,\mathcal{L}_{new}) > score(T,\mathcal{L}_{old})$.
If no such $\mathcal{L}_{new}$ exists then
the sought after partition $\mathcal{L}$ is $\mathcal{L}_{old}$ and
we are done. Otherwise put $\mathcal{L}_{old}=\mathcal{L}_{new}$ and repeat.\\ 

\noindent Note that the Heuristic move is guaranteed to terminate in 
polynomial time and that it will never return $\{L\}$ as the final partition 
\cite{JanssonEtAl2006}\footnote{To 
avoid finishing with 
$\{L\}$ we allow in the first, and only the first, iteration an operation
to be applied as long as this does not decrease the score. After this 
operations 
are only permitted if they strictly increase the score.}. Also note that the 
Heuristic move can generate
partitions with 4 or more blocks (as long as this raises the score).
This is in contrast to its analog in
\cite[page 1118]{JanssonEtAl2006} where the number of blocks in a partition
is restricted to 3. Arguably somewhat unassuming looking this restriction
to 3 blocks guaranteed that for any triplet set $T$ a level-1 network could be 
constructed which was consistent with a fraction $5/12$ of the triplets in $T$
\cite{JanssonEtAl2006}.
Interestingly - and although the system of inequalities underpinning the 
5/12 result also holds in the case when there is no restriction on the
number of blocks, as in this case - a simple level-1 network (the construction 
of which is the purpose of the second phase of \textsc{Lev1athan}) is in the 
worst case consistent with no more than $\approx 1/3$ of the triplets
in a given set. An example in point is a partition where each 
block is of size one. For such a partition the guaranteed lower bound
in the worst-case tends to 1/3. So removing the restriction to 3 blocks can
theoretically undermine the 5/12 lower bound of \cite{JanssonEtAl2006}.
However, and as suggested by examples, the
dropping of the 3-block restriction allows in practice the construction of 
a wider range of networks (and network topologies) that are consistent with a 
higher percentage of triplets (see the supplementary data section of \cite{lev1athandownload}).
If the user nevertheless requires the 5/12 lower bound to be mathematically guaranteed,
then this can be ensured by limiting the number of blocks to at most 9.

\section{Constructing Simple Level-1 Networks}
\label{sec:simple}

In this section, we turn our attention to  
the second phase of \textsc{Lev1athan} which is concerned with 
constructing a simple level-1 network from a triplet set $T$ such that
the number of triplets in $T$ consistent with that network is maximized.
Since this optimization problem
is NP-hard~\cite{JanssonEtAl2006} in general, we propose to do
this heuristically.
(We remark that for
small instances \textsc{Lev1athan} will also compare the
heuristically computed simple level-1 network with an optimal \emph{tree} computed using Wu's exponential-time
algorithm \cite{Wu2004}. If the tree is consistent with at least as many triplets as the simple level-1
network then \textsc{Lev1athan} will return the tree, parsimony again being the motivation
for this. We will not discuss this step further). Once again we emphasise that if \textsc{Lev1athan} chooses
for an Aho or JNS move when partitioning the leaf set of a triplet
set, then the algorithms described in this section will not
be used by \textsc{Lev1athan}.

To be able to describe our heuristic we need to generalize our notion
of consistency which we will do next. 
Suppose that $T$ is a triplet set and that
$\mathcal L$ is a partition of the leaf set $L=L(T)$ of $T$. Suppose 
also that $a,b,c\in L$ are distinct elements in $L$ and that
$\mathcal N$ is a network with leaf set $\mathcal L$. Then 
we say that $ab|c \in T$ is \emph{consistent}
with $\mathcal{N}$ if there exist distinct sets
$L_a,L_b,L_c\in L(\mathcal{N})$ 
with ${a\in L_a,b\in L_b,c\in L_c}$
such that $L_aL_b|L_c$ is consistent with $\mathcal{N}$. Furthermore, we 
say that~$ab|c$ is \emph{inconsistent}
with~$\mathcal{N}$ if~$ab|c$ is not consistent with~$\mathcal{N}$ 
but there do exist distinct sets $L_a,L_b,L_c\in
L(\mathcal{N})$ with $a\in L_a,b\in L_b$, and $c\in L_c$. Note that,
with this slightly generalized definition, it is possible that a triplet in $t \in T$ is
neither consistent nor inconsistent with $\mathcal{N}$ i.e. when $t$ contains two or more
leaves that lie in the same block of $\mathcal{L}$.

We are now in the position to briefly outline the second phase of 
our heuristic. 
Suppose $T$ is a triplet set and $\mathcal L$ is a partition of $L(T)$.
Then if the cardinality of $\mathcal{L}$ is moderate (by default: at most 12) we compute an exact 
optimal solution in exponential time. This exact algorithm is 
described in Section~\ref{sec:simple:exact}. If the cardinality
of~$\mathcal{L}$ is too big to compute an exact optimal solution, 
we use the greedy algorithm described in
Section~\ref{sec:simple:greedy}. 
Note that, although stated in terms of a triplet set $T$ and a
partition $\mathcal{L}$ of $L(T)$ (i.e. the partition
chosen by the previous step of \textsc{Lev1athan}), both 
algorithms can also be used for general leaf- and triplet sets: for
an arbitrary set of triplets $T'$ the partition of $L(T')$ consisting of only singletons can be taken
as $\mathcal{L}$.



\subsection{Exact Algorithm}\label{sec:simple:exact}

Van Iersel et al.~\cite{reflections} proposed an exponential-time 
exact algorithm to find a level-1 network consistent
with a maximum number of triplets of a given set $T$ in~$O(m4^n)$ time 
where $m=|T|$  and $n=|L(T)|$. This section describes how this
running time can be improved to $O(nm2^n)$ if an algorithm searches only 
for simple level-1 networks. To describe such an 
 algorithm we require two more concepts that concern
phylogenetic trees. A phylogenetic tree is said to be a 
\emph{caterpillar} if the parents of the leaves form a directed path. 
We say that a caterpillar~$C$ \emph{ends} in leaf $r$ if the sibling 
of $r$ is also a leaf of $C$ (each caterpillar thus ends in
exactly two leaves). For example the phylogenetic tree depicted in
Figure~\ref{symdiff}(a) is a caterpillar that ends in leaves $x$ and $y$.

Our algorithm, presented in the form of
pseudo-code in Algorithm~\ref{alg:SL1-1}, consists of 2 steps and
takes as input a triplet set $T$  and 
 a partition $\mathcal L$ of the leaf set $L=L(T)$ of $T$. It 
returns an optimal (in a well defined sense) simple level-1 network on $\mathcal L$. 
It works by
essentially trying each set $L_r\in \mathcal{L}$ as the leaf below the 
reticulation of the simple level-1 network to be constructed. More
precisely and with $T$ and $\mathcal L$ as
above, the first of its 2 steps is as follows.
For each set $L_r\in \mathcal{L}$ we do the following. 
For each subset~$Z\subseteq \mathcal{L}$ with $L_r\in Z$,
we first compute an optimal caterpillar~$C_Z$ ending in~$L_r$
in the sense that the number of triplets in $T$ consistent with
the caterpillar $C_Z$ is maximal. To achieve this, note
that the caterpillar $C_Z$ with $|Z|=2$ 
consists of a root and 2 distinct leaves each of which is the head of an 
arc starting at the root (lines 2-4). To 
find the other caterpillars, we loop through 
all subsets $Z\subseteq \mathcal{L}$ with $L_r\in Z$ from
small to large, starting at the subsets of size three. 
For each such subset of $\mathcal{L}$,  we loop through all elements $L'\in
Z\setminus\{L_r\}$ and create a candidate caterpillar by 
creating a new root and two arcs leaving the root: one
to~$L'$ and one to the root of~$C_{Z\setminus\{L'\}}$ (lines 5-9). 
Among all candidate caterpillars, we then 
select a caterpillar 
that is consistent with a maximum number of triplets in $T$. 

\begin{figure}[H]
\centering
      \includegraphics[scale=.7]{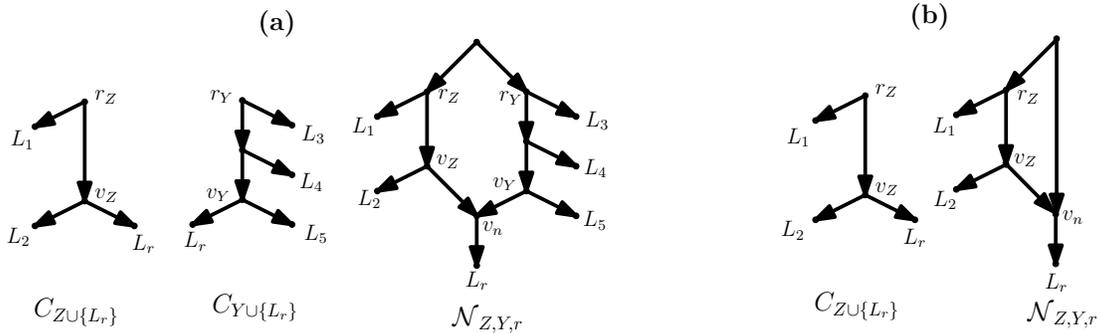}
      \caption{(a) Construction of~$\mathcal N_{Z,Y,r}$ from
$C_{Z\cup\{L_r\}}$ and~$C_{Y\cup\{L_r\}}$ if~$Z,Y\neq\emptyset$.
(b) Construction of~$\mathcal N_{Z,Y,r}$ from~$C_{Z\cup\{L_r\}}$ 
if~$Y=\emptyset$.}
      \label{fig:exactalgorithm12}
\end{figure}

Once all the caterpillars have been constructed, an optimal 
simple level-1 network is found in the second step as follows. We loop
through all bipartitions~$\{Z,Y\}$ of~$\mathcal{L}\setminus\{L_r\}$. 
First suppose that~$Z$ and~$Y$ are both nonempty.
Then we combine the caterpillars~$C_{Z\cup\{L_r\}}$ 
and~$C_{Y\cup\{L_r\}}$ into a candidate simple level-1
network~$N_{Z,Y,r}$ as follows. Let~$a_Z=(v_Z,L_r)$ be 
the arc entering~$L_r$ in~$C_{Z\cup\{L_r\}}$,
let~$a_Y=(v_Y,L_r)$ be the arc entering~$L_r$ in~$C_{Y\cup\{L_r\}}$
and let~$r_Z,r_Y$ denote the roots
of~$C_{Z\cup\{L_r\}}$ and~$C_{Y\cup\{L_r\}}$ respectively. We add 
a new root~$r_n$ and a new vertex~$v_n$ and replace
arcs~$a_Z$ and~$a_Y$ by new arcs~$(r_n,r_Z),(r_n,r_Y),(v_Z,v_n)
,(v_Y,v_n)$ and~$(v_n,L_r)$ (lines 10-14). See
Figure~\ref{fig:exactalgorithm12}(a) for an example of this construction. 
Now suppose that~$Y=\emptyset$. Let
again~$a_Z=(v_Z,L_r)$ be the arc entering~$L_r$ in~$C_{Z\cup\{L_r\}}$ 
and let~$r_Z$ be the root of~$C_{Z\cup\{L_r\}}$.
We add a new root~$r_n$ and a new reticulation~$v_n$ and replace 
arc~$a_Z$ by arcs~$(r_n,r_Z),(v_Z,v_n),(r_n,v_n)$
and~$(v_n,L_r)$. This leads to the candidate network~$\mathcal N_{Z,Y,r}$. 
See Figure~\ref{fig:exactalgorithm12}(b) for an example
of the construction in this case. Finally, we select a network 
$\mathcal N$ consistent with a maximum number of input triplets,
over all constructed candidates and return that network.

\begin{algorithm}[H]\caption{Exact Simple Level-1 Network 
Construction}\label{alg:SL1-1}
\begin{algorithmic} [1]
\REQUIRE: Triplet set $T$ and a partition $\mathcal L$ of the leaf set 
$L(T)$  of $T$. \\
\ENSURE: Simple level-1 network $\mathcal{N}$ with leaf set $\mathcal L$
consistent with a maximum number of triplets in $T$.
\FOR {$r=1\ldots q:=|\mathcal L|$} %
\FOR {$L_x\in\mathcal{L}\setminus\{L_r\}$} %
\STATE $Z\leftarrow \{L_r,L_x\}$ %
\STATE $C_Z$ is the caterpillar consisting of a 
root, the vertices $L_r$ and $L_x$, and 2 arcs both starting at
the root and one ending in $L_r$ and the other in $L_x$. %
\ENDFOR %
\FOR {$c=3,\ldots,q$} %
\FOR {each $Z\subseteq\mathcal{L}$ with~$L_r\in Z$ and~$|Z|=c$} %
\FOR {$L'\in Z\setminus\{L_r\}$} %
\STATE $C_Z^{L'}$ is the caterpillar
consisting of the caterpillar~$C_{Z\setminus\{L'\}}$, the
leaf~$L'$, a new vertex (the root of $C_Z^{L'}$), and two 
new arcs both starting
at that vertex and one ending in ~$L'$ and  the other in the root 
of~$C_{Z\setminus\{L'\}}$ %
\ENDFOR %
\STATE $C_Z$ is the caterpillar $C_Z^{L'}$ that is consistent with a 
maximum number of input triplets over all~$L'$ %
\ENDFOR %
\ENDFOR %
\FOR {all bipartitions~$\{Z,Y\}$ of~$\mathcal{L}\setminus L_r$} %
\IF {$Z,Y\neq\emptyset$} %
\STATE combine caterpillars~$C_{Z\cup\{L_r\}}$ and~$C_{Y\cup\{L_r\}}$ 
into a candidate network~$\mathcal{N}_{Z,Y,r}$ as in
 Figure~\ref{fig:exactalgorithm12}(a).
\ENDIF %
\IF {$Y=\emptyset$} %
\STATE transform caterpillar~$C_{Z\cup\{L_r\}}$ into a candidate 
network~$\mathcal{N}_{Z,Y,r}$ as in Figure~\ref{fig:exactalgorithm12}(b).
\ENDIF %
\ENDFOR %
\ENDFOR %
\STATE \textbf{return} a network~$\mathcal{N}$ that is consistent with a 
maximum number of input triplets, over all
constructed candidates~$\mathcal{N}_{Z,Y,r}$ %
\end{algorithmic}
\end{algorithm}

A straightforward analysis of the above algorithm implies the
following result.

\begin{theorem}
Given a triplet set~$T$ with $m=|T|$ and $n=|L(T)|$, 
a simple level-1 network consistent with a maximum
number of triplets from~$T$ can be constructed in~$O(nm2^n)$ time.
\end{theorem}

We now turn our attention to presenting our greedy algorithm
which follows the same philosophy as the previous algorithm i.e.
try each set in $\mathcal{L}$ as the leaf below the reticulation
of a simple level-1 network to be constructed.

\subsection{Greedy Algorithm}\label{sec:simple:greedy}

With $T$ and $\mathcal L$ as above, we next give the details of our greedy algorithm,
which we present in
the form of pseudo-code in Algorithm~\ref{alg:SL1-2}. 
In the context of this, it should be noted that a similar strategy was shown 
to perform particularly well in an algorithm by Snir
and Rao for the construction of phylogenetic trees from
 triplets~\cite{SnirRao2006}.

For each set $L_r\in \mathcal L$, we first create an initial
network~$\mathcal{N}$ with three vertices~$r,t$ and~$L_r$ and three 
arcs~$(r,t),(r,t)$ and~$(t,L_r)$ (lines 2-4), where the arc $(r,t)$
occurs twice. To this network 
we then add the
other elements from~$\mathcal{L}$ in a greedy fashion and each time 
renaming the resulting network ${\mathcal N}$. To decide which 
element $L_i\in\mathcal{L}\setminus L(\mathcal{N})$ 
to insert first, and where to insert it i.e.
into which non-trivial arc~$a$ of~$\mathcal{N}$, we use a
score function $s(L_i,a)$. To present this score function suppose
that $u$ and $v$ are vertices of ${\mathcal N}$ such that $a=(u,v)$
is a non-trivial arc of ${\mathcal N}$ and that 
$L_i\in\mathcal{L}\setminus L(\mathcal{N})$.
Let~$\mathcal{N}(L_i,a)$ denote the network obtained
from~$\mathcal{N}$ by removing arc~$a$ and adding two 
vertices~$L_i$ and~$w$ and three arcs~$(u,w),(w,v)$
and~$(w,L_i)$ to $\mathcal N$. Then the score 
$s(L_i,a)$ is equal to the number of 
triplets $t\in T$ with $L(t)\cap L_i\not=\emptyset$ 
that are consistent
with~$\mathcal{N}(L_i,a)$ minus the number of triplets 
$t\in T$ with $L(t)\cap L_i\not=\emptyset$ that are inconsistent
with~$\mathcal{N}(L_i,a)$. In other words,
\begin{eqnarray*}
s(L_i,a)&=&|\{t\in T: t \mbox{ is consistent with } 
\mathcal{N}(L_i,a)\mbox{ and } L(t)\cap L_i\not=\emptyset  \}|\\
&-& |\{t\in T: t \mbox{ is inconsistent with } 
\mathcal{N}(L_i,a)\mbox{ and } L(t)\cap L_i\not=\emptyset \}|
\end{eqnarray*}
Note that the definition of $s(L_i,a)$ does not consider triplets $t$ for which $L(t)$ 
contains at least one element in $\mathcal{L}$ that has not yet been added to the network. Also,
the definition does not consider triplets $t$ that are neither consistent nor inconsistent with $\mathcal{N}(L_i,a)$.
(This is because the role of such triplets in the final network is entirely determined by the choice of $\mathcal{L}$
and choices made in later recursive steps).

Suppose $L_i^*\in \mathcal L\setminus L(\mathcal{N})$ and 
$a^*\in A(\mathcal N)$ are such that $s(L_i,a)$ is maximized. Then 
we construct a simple level-1 network~$\mathcal{N}(L_i^*,a^*)$ and insert the
remaining leaves into this network by the same method (lines 5-8). 
Finally and by searching through all constructed simple level-1 networks
$\mathcal N_r$ we return that the network $\mathcal N$ that maximizes the
number of triplets in $T$ it is consistent with (lines 9-10). 

The algorithm can thus be summarised as follows.
\begin{algorithm}[H]\caption{Greedy Simple Level-1 Network 
Construction}\label{alg:SL1-2}
\begin{algorithmic} [1]
\REQUIRE: Triplet set $T$ and a partition $\mathcal L$ of the leaf set 
$L(T)$  of $T$. \\
\ENSURE: Simple level-1 network $\mathcal{N}$ on $\mathcal L$ (that heuristically
optimises the number of triplets consistent with $T$).
\FOR {$r=1\ldots q$} %
\STATE $\mathcal{V}_r\leftarrow \{r,t,L_r\}$ %
\STATE $\mathcal{A}_r\leftarrow \{(r,t),(r,t),(t,L_r)\}$ %
\STATE $\mathcal{N}_r\leftarrow (\mathcal{V}_r,\mathcal{A}_r)$ %
\WHILE {$L(\mathcal{N}_r)\neq\mathcal{L}$} %
\STATE compute~$s(L_i,a)$ for 
each~$L_i\in\mathcal{L}\setminus L(\mathcal{N}_r)$ and each 
nontrivial arc~$a$ of~$\mathcal{N}_r$ %
\STATE find~$L_i^*,a^*$ maximising~$s(L_i,a)$ %
\STATE $\mathcal{N}_r\leftarrow\mathcal{N}(L_i^*,a^*)$ %
\ENDWHILE %
\STATE let~$f(\mathcal{N}_r)$ be the number of triplets 
from~$T$ consistent with~$\mathcal{N}_r$ %
\ENDFOR %
\STATE let~$\mathcal{N}$ maximise~$f(\mathcal{N}_r)$ over all~$r=1\ldots q$.
\STATE \textbf{return} $\mathcal{N}$ %
\end{algorithmic}
\end{algorithm}
%

%
\section{Experiments}
\label{sec:experiments}

To better understand the behavior of \textsc{Lev1athan}, we 
performed a biologically-motivated simulation study using triplet sets 
consistent with level-1 networks of different size and
complexity and various amounts of missing data (experiment 1) and of 
noise (experiment 2). To ensure not only variability
of the input triplet sets for \textsc{Lev1athan} but also consistency 
with a level-1 network, we used a novel algorithm for random
level-1 network generation. After giving a general
outline of our simulation study in the next section,
we describe this algorithm in Section~\ref{sec:simple:networksgeneration}. 
To model missing data and noise, rather then using
the triplet set $T(\mathcal M)$ induced by such a network $\mathcal M$
we used a triplet set $T_{\epsilon}(\mathcal M)$ as input for 
\textsc{Lev1athan} where $\epsilon$ is a parameter that governs the amount of 
missing data/noise in $T(\mathcal M)$. 
Details on the precise definition of $T_{\epsilon}(\mathcal M)$
will be given in the next section.
Using a range of measures which we describe in 
Section~\ref{sec:simple:definitionsmeasures} we present the
outcomes of our study in Section~\ref{sec:simple:missingdata} in case of 
missing data and in Section~\ref{sec:simple:noiseexp} in case of noise. \\

\subsection{General outline of our simulation 
study}~\label{sec:simple:simulations}
Our simulation study consists of two experiments each of which 
is motivated by a situation one might encounter 
in a triplet based phylogenetic network analysis. The full results of (and inputs to)
both experiments are available in the supplementary data section of \cite{lev1athandownload}. The first experiment 
(Section \ref{sec:simple:missingdata} - missing data) deals with the situation 
that only some of the triplets induced by some unknown level-1 
network $\mathcal{M}$ are given. 
This phenomena is modeled by setting $T_{\epsilon}(\mathcal M)$ to be a 
randomly selected
subset of $T(\mathcal{M})$. 
The second experiment (Section \ref{sec:simple:noiseexp} - noise) addresses 
the situation that the confidence level one might have in the 
triplets generated 
in a phylogenetic analysis might vary. In our experiment 
this is modeled by adding
noise to $T(\mathcal{M})$. Put differently, we essentially construct 
$T_{\epsilon}(\mathcal M)$ by randomly 
selecting triplets from $T(\mathcal{M})$ and 
for each such selected triplet $t$ randomly ``flipping'' it to one of the two
other possible triplet topologies on $L(t)$.

For both simulation experiments, the general outline 
is as follows. We first
choose some level-1 network $subNet$ as the ``seed'' 
for our random level-1 generator algorithm. For the generated 
level-1 network $\mathcal{M}$, we then constructed the triplet set
$T_{\epsilon}(\mathcal{M})$ from
$T(\mathcal{M})$ and use $T_{\epsilon}(\mathcal{M})$ 
as input for \textsc{Lev1athan}. The
level-1 network $\mathcal{N}$ generated by \textsc{Lev1athan}
we then compared against $\mathcal{M}$ with regards to 
topology and also the four measures described 
in Section~\ref{sec:simple:definitionsmeasures}. 
In each experiment we used a total of 110 randomly generated level-1 networks 
with leaf set size ranging between $22$ and $115$ and number of reticulations
ranging between $1$ and $10$. 
The running time of \textsc{Lev1athan} on a standard desktop computer 
ranged from 2-3 seconds for the
simplest networks to 30 seconds for the most complex ones. 

\subsection{Generating random level-1 
networks}\label{sec:simple:networksgeneration}

A survey of the literature suggested the NetGen \cite{netgen} program to
be the only available approach for systematically generating 
networks. Whilst NetGen addresses the issues of size 
(i.e. number of vertices) and network complexity one would encounter
when manually constructing networks, one of its main drawbacks  
is the lack of guarantee that the generated network is level-1
(although some happen to be level-1 networks). One way to overcome this 
problem is to hand pick suitable networks from a (relatively small) subset of 
generated networks. Alternatively the list $L_{NetGen}$ 
of networks generated by
NetGen could be post-processed by manually
removing a sufficient number of reticulations from each
network in $L_{NetGen}$ to obtain a level-1 network.
A further and maybe more important drawback of NetGen is that 
the structure of a gall in a generated network is rather simple in the
sense that its size is 4. We therefore developed our own algorithm for
generating level-1 networks. This algorithm is implemented in Java 
and freely available for download from \cite{lev1athandownload}. 
We next describe this algorithm.

\begin{figure}[H]
\centering
\includegraphics[scale=.6]{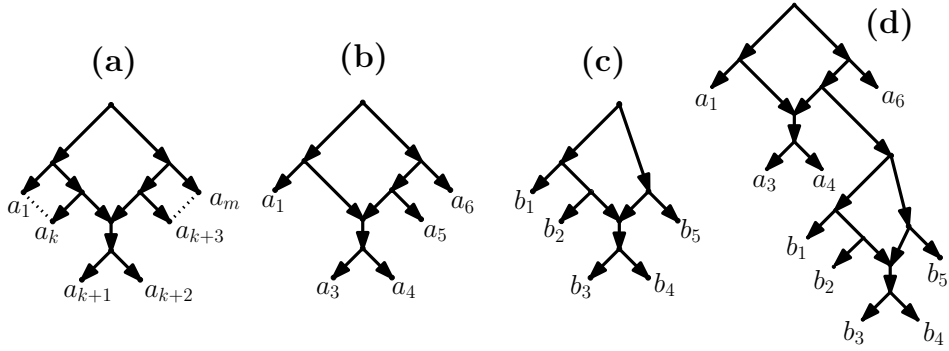}
\caption{\label{fig:gen1}
The figure illustrates the generation of a level-1 network 
via our level-1 network generator algorithm (c.\,f.\,Algorithm~\ref{alg:gn}). 
The initial network $subNet$ 
is depicted in (a). In the discussed example, $m=6$ for the network
pictured in (a).  The networks 
presented in (b) and (c) are obtained by randomly removing 
vertices from two instances of the network in (a). The network depicted in
(d) is the network $\mathcal{M}$ 
generated by replacing leaf 
$a_5$ in the network pictured in (b) with the network shown in (c). } 
\end{figure}

Our algorithm for generating level-1 networks takes as 
input a level-1 network $subNet$ on a fixed number $m$ of leaves plus a 
positive integer $n$ and outputs a level-1 network $\mathcal M$
on a larger leaf set. A pseudo-code form of the algorithm is 
presented in Algorithm \ref{alg:gn}.
\begin{algorithm}[H] \caption{Level-1 network generator} \label{alg:gn}
\begin{algorithmic}[1]
\REQUIRE : A simple level-1 network $subNet$  and a positive  integer $n$.   \\
\ENSURE: Level-1 network $\mathcal{M}$
\STATE $ \Sigma \leftarrow \emptyset$
\STATE $ \mathcal{M} \leftarrow $ empty graph
\FOR {$i=1\ldots n$}
	\STATE ${N_i}\leftarrow$ generate instance of $subNet$
	\STATE ${\Sigma}\leftarrow {\Sigma} \cup N_i$	
\ENDFOR 
\FORALL {${N_i} \in {\Sigma}$}
\STATE relabel the leaves in $N_i$ such that no two networks in $\Sigma$
have the same leaf set,
	\STATE randomly choose some integer $l$ in $ \{0,1,\ldots,
\lceil |V(subNet)|/4\rceil \}$
	\STATE randomly delete $l$ vertices from $N_i$
\ENDFOR 
\STATE $\mathcal{M}\leftarrow N_1$
\FOR {$i=2 \ldots n$}
	\STATE randomly choose a leaf of $\mathcal{M}$ and replace it with
the network $N_i$
\ENDFOR 
\STATE $ {p} \leftarrow $ number of leaves in $\mathcal{M}$
\STATE randomly choose some integer $l'$ in 
$\{0,1,\ldots, \left\lceil p/2 \right\rceil \}$
\STATE randomly delete $l'$ leaves and cherries from $\mathcal{M}$
where, as usual, a cherry of a network $\mathcal N'$ 
is a set of leaves $a$ and $b$ of $\mathcal N'$
such that the parent of $a$ is also the parent of $b$.
\STATE \textbf{return} $\mathcal{M}$
\end{algorithmic}
\end{algorithm}
We remark that the size of the gall $C$ in $subNet$  clearly
influences the variability of networks generated by this algorithm. 
Also, we remark that for the purpose of the discussed simulation
study $subNet$ was the level-1 network depicted in Figure~\ref{fig:gen1}(a). 

Algorithm~\ref{alg:gn} starts by generating 
$N_1 \ldots N_n$ instances 
of $subNet$ and storing them in set $\Sigma$ (lines 3 -- 5).
Next (lines 6-9), for each $N_i \in \Sigma$ we first  randomly chose
 an integer $l$ with $0 \leq l \leq \lceil |V(subNet)|/4\rceil$ 
and then randomly delete $l$ vertices 
$v_q$, $1 \leq q \leq l$ from $N_i$ making sure that no deleted vertex 
is the root or a leaf of $N_i$. 
If $v_q$ is a reticulation then we randomly choose one of 
the parents  of $v_q$, say $p_q$, and add a new arc from $p_q$ 
to the unique child of $v_q$. 
For any other deleted vertex $v_q$ we reconnect its unique parent with one 
of its children. The child to be reconnected is selected randomly, 
but a child that is not a leaf is always preferred over a child that is
a leaf. We initialize our output level-1 network $\mathcal{M}$ with the 
network $N_1$ (line 10) and
then iterate over $N_i$, where $i = 2 \ldots n$. At 
each iteration we randomly 
select a leaf from network $\mathcal{M}$ and replace it 
by $N_i$ (line 12) yielding
a new level-1 network $\mathcal{M}$. Finally, we randomly remove 
a random number of leaves and cherries from $\mathcal{M}$ (lines 13-15)
and then return the resulting network which we also call $\mathcal M$. 
We conclude the description of  Algorithm \ref{alg:gn} by making the
trivial observation  that 
the size of $\mathcal{M}$ depends on the number $n$ of $subNets$.

To illustrate the inner workings of the level-1 network generator algorithm 
suppose $subNet$ is the level-1 network  with leaf set $\{a_1,\ldots, a_m\}$
depicted in Figure~\ref{fig:gen1}(a) with $m=6$ and suppose that $n=2$. 
Then we first generate 
$2$ instances of that network. The deletion of vertices  
from each of that networks as described in line 9 of Algorithm~\ref{alg:gn} 
results in the networks depicted in Figures~\ref{fig:gen1}(b) and 
\ref{fig:gen1}(c). Next,  leaf $a_5$ in 
the network depicted in Figure~\ref{fig:gen1}(b) is 
replaced with the network pictured in Figure~\ref{fig:gen1}(c). 
The resulting network $\mathcal{M}$ is depicted in Figure~\ref{fig:gen1}(d). 
Note that due to the small number of leaves of {\em subNet}, the operation of 
removing random leaves and cherries from that network 
(lines 13-15) is not executed.

\subsection{Measures}\label{sec:simple:definitionsmeasures}

Reflecting the fact that our simulation study is aimed at assessing
the reconstructive power of \textsc{Lev1athan} by measuring 
the similarity between a level-1 network $\mathcal M$
and the level-1 network $\mathcal N_{{\mathcal M}, \epsilon}$ generated 
by \textsc{Lev1athan} when given $T_{\epsilon}(\mathcal M)$, we considered 4 different measures. 
To be able to describe these measures, we need to fix some notation first. 
For the remainder of this section and unless stated otherwise, 
let $\mathcal M$ be a level-1 network 
and let 
$\mathcal{N}_{\mathcal M}:=\mathcal{N}_{\mathcal M,\epsilon}$ 
denote the network that \textsc{Lev1athan} generates when given 
$T_{\epsilon}(\mathcal M)$ as input.

The first measure is the \emph{network-network triplet consistency measure}.
For $\mathcal M$ and $\mathcal{N}_{\mathcal M}$ we define the
network-network triplet consistency measure
$C(\mathcal{M}, T_{\epsilon}(\mathcal M), \mathcal{N}_{\mathcal M})$ 
as the quantity
\[
C(\mathcal{M}, T_{\epsilon}(\mathcal M), \mathcal{N}_{\mathcal M} )  
=\frac{|T(\mathcal{M}) \cap T_{\epsilon}(\mathcal{M}) 
\cap T(\mathcal{N}_{\mathcal M})|}{|T(\mathcal{M}) 
\cap T_{\epsilon}(\mathcal M)|}.
\]
Loosely speaking, 
$C(\mathcal{M}, T_{\epsilon}(\mathcal M), \mathcal{N}_{\mathcal M}) $  
is the proportion of ``definitely correct'' triplets (i.e. 
$T(\mathcal{M}) \cap T_{\epsilon}(\mathcal{M})$) that are consistent 
with $\mathcal{N}_{\mathcal M}$. Thus 
$C(\mathcal{M}, T_{\epsilon}(\mathcal M), \mathcal{N}_{\mathcal M} )$ is  
always a real number between $0$ and $1$. 
A variant of this, the {\em triplet-network triplet consistency measure}
$C(T, \mathcal P)$, pays less heed to the origin or accuracy of the input triplets
and is defined for an arbitrary 
triplet set $T$ and a phylogenetic network $\mathcal P$ by putting
$$
C(T, \mathcal P)=\frac{|T \cap T(\mathcal{P})|}{|T|}.
$$ 
In other words,
$C(T, \mathcal P)$ is the fraction of triplets 
in $T$ that is consistent with the network $\mathcal P$. Note that 
this measure is different from the triplets distance introduced in \cite{Cardona2009b}.

The third of our triplet based measures is inspired by the 
quartet distance for unrooted phylogenetic trees \cite{Bryant2000} and is
called the \emph{network-network symmetric difference}. For
$\mathcal M$ and $\mathcal N_{\mathcal M}$
the network- network symmetric difference
$S(\mathcal{M}, \mathcal{N}_{\mathcal M} )$ between $\mathcal M$ and 
$\mathcal N_{\mathcal M}$ is defined as the quantity 
\[
S(\mathcal{M}, \mathcal{N}_{\mathcal M} )  = 
|T(\mathcal{M}) \Delta T(\mathcal{N}_{\mathcal M})|.
\]
$S(\mathcal{M}, \mathcal{N}_{\mathcal M} )$ is thus the number of triplets that appear in $T(\mathcal{M})$ but not 
in $T(\mathcal N_{\mathcal M})$, or vice-versa. Note that in this measure $\mathcal M$ and $\mathcal N_{\mathcal M}$
are compared with regards to their induced triplet sets, while 
$\mathcal N_{\mathcal M}$ was generated in response to the set $T_{\epsilon}(\mathcal{M})$
derived from $T(\mathcal{M})$.
As already noted above for the network-network triplet 
consistency measure, the network-network symmetric difference
measure also suggests a natural variant of it, $S(T, \mathcal N)$, defined for an arbitrary 
triplet set $T$ and a network $\mathcal N$ by putting
$S(T, \mathcal{N}) = | T \Delta T(\mathcal{N})|$. 

Regarding the $S(T, \mathcal{N})$ and $C(T, \mathcal N)$ measures
it should be noted that the former might be more powerful 
than the latter. To see why consider again the example of the
triplet set $T(\mathcal N)$ induced by the caterpillar $\mathcal N$ on
$X=\{x,y,z,g,f\}$ depicted in Figure~\ref{symdiff}(a).
As already pointed out earlier, the level-1 network 
$\mathcal N'$ on $X$ depicted in Figure~\ref{symdiff}(b)
is also consistent with $T(\mathcal N)$ and no arc of 
$\mathcal N'$ can be deleted from $\mathcal N$ so that consistency 
with $T(\mathcal N)$ is preserved. 
If we take
$T(\mathcal N)=T_\epsilon(\mathcal{N})$, 
then (in the context of network-network triplet consistency) $\mathcal N$ and $\mathcal N'$ are equally good level-1 networks 
for representing $T(\mathcal N)$ since
$T(\mathcal N)\subsetneq T(\mathcal N')$. However, under the network-network symmetric
difference measure $\mathcal N$ would be preferable over $\mathcal N'$, 
because of precisely that proper subset relationship.

The final measure we considered is the $\mu$-distance $d_{\mu}$ which was
originally introduced in \cite{Cardona2007}. To define this measure
which was shown in \cite{Cardona2009a,Cardona2007} to be a metric for a certain class of
networks (i.e. \emph{tree-child} networks), which includes the class of level-1 networks 
as a subclass, we require some more notation. Suppose $\mathcal N$
is a phylogenetic network on some set $X=\{1,\ldots, n\}$, $n\geq 1$, 
and $v$ is a vertex of $\mathcal N$. Then the vector 
$\mu_{\mathcal N}(v)=(m_1(v),m_2(v),\ldots,m_n(v))$ can be associated 
to $v$ where
for all $i\in X$ the quantity $m_i(v)$ represents the number of different
paths in $\mathcal N$ from $v$ to leaf $i$. With denoting
by $\mu(\mathcal N)$ the multiset of all vectors 
$\mu_{\mathcal N}(v)$, $v$ a vertex of $\mathcal N$,
and calling $\mu(\mathcal N)$ the {\em $\mu$-representation} of $\mathcal N$, 
the $\mu$-distance
$d_{\mu}(\mathcal N_1,\mathcal N_2)$ between any two phylogenetic networks
$\mathcal N_1$ and $\mathcal N_2$ is defined as
$$
d_{\mu}(\mathcal N_1,\mathcal N_2)=|\mu(\mathcal N_1)\Delta \mu(\mathcal N_2)|
$$
where the symmetric difference operator is defined here over multisets.

Armed with these measures and our algorithm for randomly generating
level-1 networks, we are now ready to describe the results of our 
simulation study. (We note that, because the source networks used in the
simulation had no vertices of outdegree 3 or higher, and for technical reasons, the optional
post-processing phase used by \textsc{Lev1athan} was switched \emph{off} during
these simulations). Assume then from now on that $\mathcal M$ 
is a level-1 network generated by our random level-1
network generator described in Algorithm~\ref{alg:gn} 
and that the definition of the network $\mathcal N_{\mathcal M}$ is as before.
We start with presenting our results for the missing data experiment.

\subsubsection{Simulation results - missing 
data experiment}\label{sec:simple:missingdata}

Central to this experiment is the parameter $\epsilon$ 
which indicates the probability that a triplet
in $T(\mathcal{M})$ will be included in $T_{\epsilon}(\mathcal{M})$. 
The values of $\epsilon$ that we investigated were $\epsilon=1.0$ 
(i.e. all triplets in $T(\mathcal{M})$), $0.9,\ldots, 0.4$.

Modulo a well-understood exception (described below) all networks 
$\mathcal{M}$ generated via 
Algorithm \ref{alg:gn} were recovered correctly by \textsc{Lev1athan}
when $\epsilon=1.0$. This is a consequence of the fact that \textsc{Lev1athan}
prioritises JNS moves and that a level-1 network $\mathcal{M}$ is completely
defined by $T(\mathcal{M})$ up to, but not including, galls of size 4~\cite{GambetteHuber2009}.  
This is a drawback of any triplet based phylogenetic network approach since such approaches 
have to make a choice between the three galls on a set $X=$\{$a,b,c$\} that are all consistent with $T=\{ab|c, a|bc\}$. 
This inability to distinguish between the topologies of size-4
galls does not, of course, affect the triplet-based measures, and for $\epsilon=1.0$
we had indeed in all cases that $S(\mathcal{M},\mathcal{N}_{\mathcal M})=0$ and
$C(\mathcal{M}, T_{\epsilon}(\mathcal M), \mathcal{N}_{\mathcal M} )=1$. When
\textsc{Lev1athan} correctly guessed the topologies of all size-4 galls in a network $\mathcal M$
we additionally had $d_{\mu}(\mathcal{M},\mathcal{N}_{\mathcal M})=0$, but this
value became non-zero when the guess was incorrect.

\begin{figure}[H]
\centering
      \includegraphics[scale=.5]{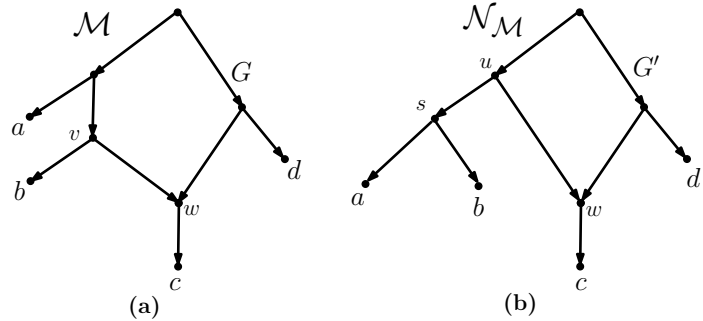}
      \caption{The figure illustrates typical gall damage. (a) The level-1 
network $\mathcal M$ on $X=\{a,\ldots, d\}$ with a gall $G$ of size 5. 
(b) The level-1 network $\mathcal N_{\mathcal M}$ on the same set with the
erroneous reconstruction of $G$ in terms of the gall $G$'.}
\label{galldamage}
\end{figure}

For all other values of $\epsilon$ and all networks $\mathcal M$,
we frequently observed that 
- although similar when inspected visually - some of 
the galls from $\mathcal{M}$
were not reconstructed correctly in the generated level-1 network
$\mathcal N_{\mathcal M}$, in the sense that the size 
of a gall in $\mathcal{N}_{\mathcal M}$ was smaller than 
in the corresponding gall in $\mathcal{M}$ (e.g. a size 5 gall in
$\mathcal M$ became size 4 gall in $\mathcal{N}_{\mathcal M}$, 
see Figure~\ref{galldamage}). Although this observation is clearly dependent 
on the specific value used for $\epsilon$, it generally meant that one 
or more vertices had been pushed out of a gall in $\mathcal M$ to its sides, 
causing what we will call \textit{typical gall damage} for $\mathcal M$. In turn this means
that subnetworks hanging from galls are sometimes merged by \textsc{Lev1athan} into
a single subnetwork. Even in the presence of typical gall damage, however, 
\textsc{Lev1athan} sometimes (but not always) correctly inferred which 
leaf of $\mathcal M$ needed to be placed
below the reticulation of the damaged gall.
 Interestingly, we also observed that typical gall damage was rare for 
galls $G$ - even for low vales 
of $\epsilon$ - if the subnetworks 
hanging from $G$ contain many leaves. 
Expressed differently, the likelihood that a gall in $\mathcal{M}$ 
suffers typical gall damage for $\epsilon < 1.0$ is higher if the gall 
$G$ 
is closer to the leaves. The reason for this might be that $G$ 
is supported by far fewer triplets then a gall closer to the root. 

It should be noted that the exception to the former of the last two
observations is 
the galled caterpillar network \cite{ByrkaEtAl2008,JanssonEtAl2006}
which can be thought of as a natural level-1 generalization of a caterpillar.
Such networks $\mathcal M$ were correctly
reconstructed by \textsc{Lev1athan} for $\epsilon > 0.8$.

In addition, we observed for all networks $\mathcal M$ that 
when $\epsilon$ dropped from 1.0 to 0.9, the network-network triplet
 consistency between $\mathcal M$ and $\mathcal N_{\mathcal M}$ 
also drops slightly (often from 1 to a value in the range 0.95 - 0.99) 
but immediately stabilises around that value,
even for very low values of $\epsilon$. This phenomenon even 
occurs when there are very
few or even no Aho or JNS moves occurring, suggesting 
that Heuristic moves are (in this context)
good for sustaining a very high value of triplet consistency.
However, and for all values of $\epsilon$ other than $1.0$ and all
networks $\mathcal M$, we also observed that  
even when 
$C(\mathcal{M}, T_{\epsilon}(\mathcal M), \mathcal{N}_{\mathcal M} )$ 
is very close to 1, $\mathcal{M}$ is often not recovered correctly 
by \textsc{Lev1athan} from $T_{\epsilon}(\mathcal M)$. An example that illustrates
this point is
the triplet set $T(\mathcal N)$ of the level-1 network 
$\mathcal N=\mathcal N_{\mathcal M}$
depicted in Figure~\ref{galldamage}(b). This triplet set together
with the triplet $bc|a$ is the triplet set $T(\mathcal M)$ induced by
the level-1 network $\mathcal M$ depicted in Figure~\ref{galldamage}(a).
The network generated by \textsc{Lev1athan} from $T_{\epsilon}(\mathcal M)=T(\mathcal N)-\{bc|a\}$ 
is the level-1 network $\mathcal N$ and 
$C(\mathcal{M}, T_{\epsilon}(\mathcal M), \mathcal{N} )=1$.
Thus, rather than being a suitable tool for assessing \textsc{Lev1athan}'s
reconstructive power, the 
$C(\mathcal{M}, T_{\epsilon}(\mathcal M), \mathcal{N}_{\mathcal M} )$ 
measure might be
of limited use for capturing differences between networks.

For all values of $\epsilon$ and averaged over
all networks $\mathcal M$ for $\epsilon$, 
we observed an initial sharp rise
for both the network-network symmetric difference and the
$\mu$-distance
when $\epsilon$ drops from $1.0$ to $0.9$. Again
in both cases this initial rise is followed by a a much slower
growth rate although this rate seems to be higher
for the $\mu$-distance. Encouragingly, we found
instances of networks $\mathcal M$ where for $\epsilon \geq 0.8$ 
\textsc{Lev1athan} correctly inferred the missing triplet information, that is, 
correctly reconstructed $\mathcal M$ from $T_{\epsilon}(\mathcal M)$.

\subsubsection{Simulation results -  noise 
experiment}\label{sec:simple:noiseexp}


Central to this experiment is again the parameter 
$\epsilon$ which in this case is an
error probability and which we state in terms of values between 
$0$ and $1$.  
Its purpose is to introduce noise into $T(\mathcal{M})$ 
and the range we considered was  $0.00,0.01,0.02 \ldots 0.10$ and 
$0.10,0.15,\ldots,0.50$. 
More precisely, we generated
 $T_{\epsilon}(\mathcal{M})$ from $T(\mathcal{M})$ by taking each triplet 
$t \in T(\mathcal{M})$ and with
probability $\epsilon$ decide to \emph{corrupt} it, that is, replace 
$t$ in $T(\mathcal M)$ with one of the 2 other phylogenetic trees
on $L(t)$, chosen uniformly at random. 

We observed that even for very low error probability values,
the triplet set $T_{\epsilon}(\mathcal M)$ is unlikely 
to be consistent with a level-1 network.
It is thus not
surprising that we often  observed additional (erroneous) 
reticulations in networks reconstructed
 by \textsc{Lev1athan} from $T_{\epsilon}(\mathcal M)$. 
Nevertheless and based on visual inspection,
our experiment seems to indicate that even for slightly higher
values of $\epsilon$, i.\,e\, $\epsilon=0.05,0.06,\ldots,0.15$, 
and all networks $\mathcal M$, much 
 of the structure of $\mathcal{M}$ is recovered correctly 
by \textsc{Lev1athan} from $T_{\epsilon}(\mathcal M)$. Having said 
that, and in addition to the above
mentioned erroneous reticulations, typical gall damage is 
common in the generated networks.

\begin{figure}[H]
\centering
\includegraphics[scale=.5]{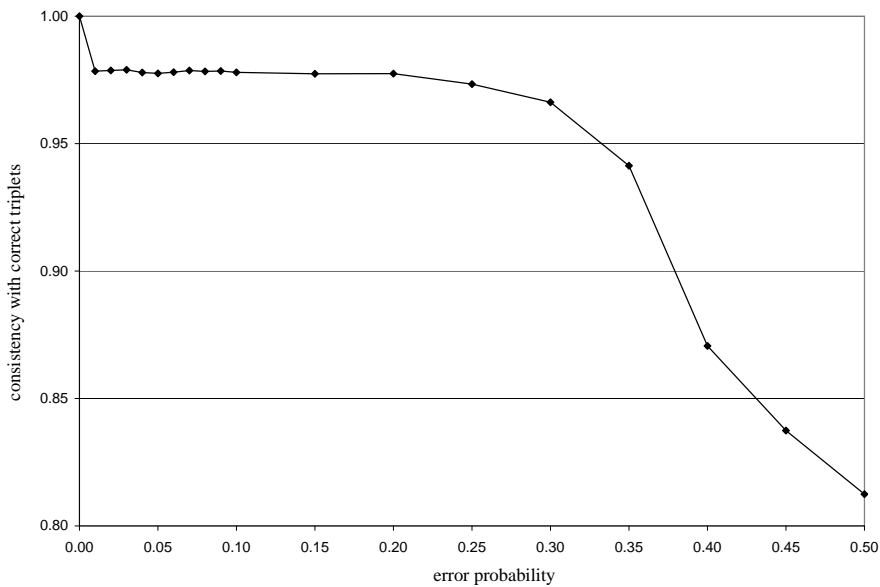} 
\caption{In the noise experiment the $C(\mathcal{M}, T_{\epsilon}(M), \mathcal{N}_{\mathcal M} )$ measure experiences an initial sharp
drop of about two percent, after which it remains relatively stable until the error probability becomes large (bigger than 0.3).}
\label{cons_drop}
\end{figure}

For all $\epsilon$ other than 0.00 and averaged over all networks $\mathcal M$ 
for $\epsilon$, we observed that, as expected, 
$C(T_{\epsilon}(\mathcal{M}), \mathcal{N}_{\mathcal M})$ 
decreased linearly with increasing error probability in the sense that   
$C(T_{\epsilon}(\mathcal{M}), \mathcal{N}_{\mathcal M}) \approx  1 - \epsilon$ 
 holds. Reassuringly (and by no means obviously) this almost linear relationship is \emph{not} obeyed by the
network-network triplet consistency measure 
$C(\mathcal{M}, T_{\epsilon}(M), \mathcal{N}_{\mathcal M} )$. 
By averaging for each error probability value over all networks $\mathcal M$,
we summarize our findings for that measure in Figure \ref{cons_drop}
in terms of plotting the error probability values for $\epsilon$ 
against the corresponding 
network-network triplet consistency values.  
 Intriguingly, there is an initial 2\% drop after which network-network 
triplet consistency remains high until error rates reach values of 
$\epsilon > 0.30$. It should be noted that
this in fact corresponds to a high level of noise in $T(\mathcal M)$
given that if $\mathcal{M}$ is a phylogenetic tree and $\epsilon=0.66$ then  
$T_{\epsilon}(\mathcal{M})$ is a completely
randomized triplet set and \emph{all} structural information 
contained in $T_{\epsilon}(\mathcal{M})$ concerning $\mathcal M$ 
has been lost.
The most likely explanation for the initial 2\% drop is probably (again)
typical gall damage since, as alluded to above, a given gall represents 
more triplets then its a gall-damaged counterpart. 
The fact that after the initial drop 
network-network triplet consistency remains high, 
is encouraging, because it shows that \textsc{Lev1athan} holds
promise for reconstructing triplets that have not been corrupted 
by noise, up to quite a high level of noise.
\noindent 

For all values of $\epsilon$ and averaged over
all networks $\mathcal M$ for $\epsilon$,
we also observed an initial sharp increase when $\epsilon$ increases 
form $0$ to $0.01$ for both the network-network symmetric difference 
and the $\mu$-difference. After this, the values for 
both measure grow very slowly 
with the growth rate for the $\mu$-distance being higher until
$\epsilon$ reaches $0.3$ when the growth rate of the network-network 
symmetric difference takes over. Intriguingly the
network-network symmetric difference seems to reach a peak at 
$\epsilon=0.4$ and then drops of again for $\epsilon=0.45$. However and in
the light of the fact that, as pointed out above, $0.3$ is already
a high error probability the later 2 observations should be treated
with caution. A possible reason for the above mentioned initial sharp rise might be that even
one corrupted triplet can potentially introduce erroneous (additional)
galls in $\mathcal N_{\mathcal M}$. Combined with the problem of typical
gall damage, this can greatly affect the
triplet sets induced by $\mathcal M$ and  $\mathcal N_{\mathcal M}$
as well as their $\mu$-representations and thus the network-network
symmetric difference and the $\mu$-distance which rely on these concepts respectively.

Interestingly we identified some networks $\mathcal{M}$
such that when $\epsilon=0.01$ (and additionally $T_{\epsilon}(M) \neq T(M)$) we nevertheless had
that $d_{\mu}(\mathcal M,\mathcal{N}_{\mathcal M})=0=S(\mathcal M,\mathcal{N}_{\mathcal M})$.
Visual inspection revealed that in such cases $\mathcal M$ was equal to  
$\mathcal{N}_{\mathcal M}$. This suggests that 
if the noise level in an input triplet set is small enough
\textsc{Lev1athan} might still be able to correctly reconstruct the level-1 network
that induced that triplet set.

\subsection{A HIV-1 virus data set}\label{sec:simple:bioexample}

To illustrate the applicability of our approach, we re-analyzed a HIV-1
virus data set that originally appeared  in 
\cite[Chapter 14]{phylogenetics-handbook}.
All but one of the sequences ({\em KAL153}) making up that data set are  
50 percent consensus sequences for  the HIV-1 M-group subtypes $A,\ldots, D$,  
$F,G,H$, and $J $ with the 
{\em KAL153} strain being thought to be a recombinant of subtypes $A$ and $B$. 
Recombinants such as {\em KAL153} can essentially be thought of as having  
arisen via the transfer and integration of genetic material from one 
evolutionary lineage into another.
The positions in an existing sequence where the foreign genetic material 
was integrated are generally called breakpoints and many approaches for 
detecting recombination aim to identify these breakpoints. 

\begin{figure}
\centering
\includegraphics[scale=.5]{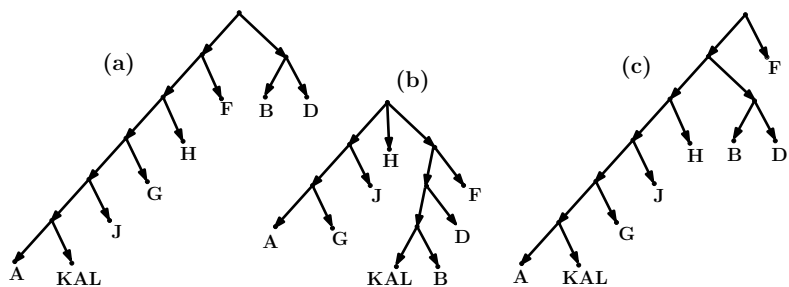} 
\caption{The phylogenetic trees for the three sub-alignments with 
the outgroup $C$ omitted in
each case. (a) sub-alignment 1-2699, (b) sub-alignment 2700- 8925, and 
(c) sub-alignment 8926-9953.  }
\label{virus-trees}
\end{figure}

For the above data set two breakpoints were identified 
(positions 2700 and 8926) in \cite[Chapter 14] {phylogenetics-handbook}.
 Furthermore, for the  three induced sub-alignments 1-2699, 2700- 8925, and 
8926-9953 the Neighbor Joining method \cite{nei2000} (with subtype $C$ 
as outgroup and the K2P model \cite{Kimura1980}) was 
used to represent the data 
in terms of arc-weighted phylogenetic trees (see  \cite[page 159]{phylogenetics-handbook} 
for a depiction of those trees). Since the resolution 
patterns for $J$ and $G$ in the first tree, $H$ and $C$ in the second, and 
$G$ and $J$ in the third tree was not clear, we recomputed those trees using 
the above settings. Reassuringly and when arc weights were ignored, 
this resulted in the same phylogenetic trees for the 
first and third sub-alignment as in the previous analysis except that the
unresolved vertex in each tree was now resolved. For the second sub-alignment,
the same tree was obtained. For the convenience of the reader, we depict 
these phylogenetic trees in Figure~\ref{virus-trees}.

As expected, the position of  the {\em KAL153} strain is the same 
in the first and third topology but different in the second topology. 
However and somewhat surprisingly the resolution order of subtypes $G$ 
and $J$ is in the first and third tree is different as well as the 
placement of subtype $F$ in the tree. Having said this, the branch 
weights that support these different resolution 
orders are very small. To therefore not allow this to overly 
influence our analysis (after all our method as well as the other two
methods that we tried out are using phylogenetic trees rather than
arc-weighted phylogenetic trees as input and therefore these different levels 
of support are not taken into account), we only used the first and the 
second and the third and second tree as respective inputs for our analysis. 

\begin{figure}
\centering
\includegraphics[scale=.6]{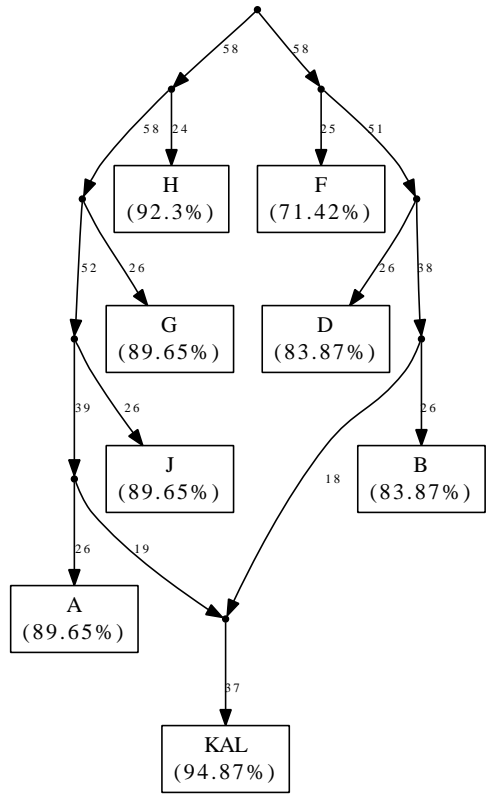} 
\caption{The optimal simple level-1 network constructed by 
\textsc{Lev1athan} for the first
and second tree in  Figure~\ref{virus-trees}. For each taxa $x$
the percentage value denotes the percentage of input triplets
containing $x$ that the network is consistent with. This can be interpreted
as how ``satisfied'' the taxa in question is with its location in the network. The
value on each arc is the {\em deletion support value} for that
arc i.e. the number of input triplets that would cease to be consistent with 
the generated network if that arc was deleted.}
\label{bestsimple-tree1-2}
\end{figure}

Interestingly the second phylogenetic tree postulates the triplet 
$FB|A$ on subtypes $A$, $F$ and $B$ whereas the first tree 
hypotheses $AF|B$. Since this conflict also interferes with the 
conflicting information for subtypes $A$, $B$ and {\em KAL153}, 
attempting to combine the triplet set generated from both phylogenetic trees 
into a general level-1 network is problematic as such a network 
would postulate 2 recombination events for this data set. 
To avoid this and at the same time identify the stronger of the two 
conflicting signals it is therefore more useful to construct (using
Algorithm~\ref{alg:SL1-1}) an optimal {\em simple} level-1 network, 
which by definition has
only one reticulation vertex. This type of 
network is depicted for the first and second tree
from  Figure~\ref{virus-trees} in Figure~\ref{bestsimple-tree1-2}. 
As expected, the network correctly identifies
the $KAL153$ strain as a recombinant of subtypes $A$ and $B$.

It should be noted that repeating this analysis for the 
second and third tree in  Figure~\ref{virus-trees}.
resulted in the same simple level-1 network as the one 
depicted in Figure~\ref{bestsimple-tree1-2}
except that the order of $G$ and $J$ was reversed.
The same holds when the optimal simple level-1 
network is computed for the respective original 
phylogenetic trees from  \cite[Chapter 14]{phylogenetics-handbook}
when ignoring arc-weights. 
Interestingly and in case of the second and the third tree,
exclusion of subgroup $F$ also resulted in a simple
level-1 network that correctly identified the 
{\em KAL153} as recombinant. However this was not the case
when this analysis was repeated for the first and second tree.

We conclude this section with remarking that although using 
different philosophies, the other two known approaches i.e. 
{\em Cluster networks} 
and {\em Galled networks} (both of which are implemented in 
Dendroscope \cite{Huson2007Dendroscope}) also had problems
with this data set, postulating between 2 and 4 recombination events 
(data not shown).

\section{Conclusions}
\label{sec:conclusions}

In this paper, we have presented a heuristic 
for constructing level-1 networks from triplet sets
which we have also implemented in Java as the freely available
program \textsc{Lev1athan} \cite{lev1athandownload}. Guided by the 
principle of triplet consistency, our heuristic aims
to optimize a well-defined objective function on triplet sets
without generating pessimistically complex networks for them. 
By running in polynomial time and always 
returning a level-1 network, it addresses several of the problems that 
frequently occur with existing network algorithms from such input sets.
Using both a biologically motivated simulation study 
and a biological data set, we have also explored \textsc{Lev1athan}'s
applicability to real biological data. 


Based on the outcomes of our simulation study, it appears that 
our heuristic is able to recover, in terms of the
triplet set induced by the generated level-1 network, a high 
percentage of the input triplets that were also present in the original
network (as opposed to triplets that were missing or that had been corrupted).
This is nota bene also true when the input 
triplet set is not dense, which (in light of the NP-hardness of 
the non-dense case) is an encouraging observation.
On the other hand (and probably not surprisingly 
as our heuristic tries to reconstruct a global structure from 
local information) it seems that it is more vulnerable to noise in an input
triplet set than missing data. The most probable reason for this is
that the former type of problem can sometimes 
be rectified via implicitly inferred triplet information, whereas the
latter type of problem has the capacity to actively mislead. Having said 
that, \textsc{Lev1athan} shows encouraging potential
if the amount of missing data is low or there is only very little
noise in the data.

In general using more of the triplets induced by a network 
as input for \textsc{Lev1athan} allows larger regions of that network
to be recovered by it. Having said that, when confronted with 
missing data or noise, even 
extremely high values of network-network triplet consistency 
(e.g. above 0.95) do not preclude non-trivial  
differences between the original network and the network generated by 
\textsc{Lev1athan}. Additionally, the noisier a input triplet set for 
our heuristic is, the greater the chance that 
the network found by it is distorted (e.g. through
the emergence of surplus galls in the generated network). To tackle
this problem \textsc{Lev1athan} has the option to label each arc of the generated
network with its \emph{deletion support value}, see Figure
\ref{bestsimple-tree1-2}. This allows the
user to explore which reticulations 
are weakly supported, and thus might be superfluous or even artefacts
of our heuristic.

Although the four triplet-based measures used in this article
appear to be very natural they only seem to be of limited 
use for capturing network 
differences in general. 
However some of them helped to identify cases where a network 
generated by \textsc{Lev1athan}
from $T_{\epsilon}(\mathcal{M})$ coincided
with the original network $\mathcal M$.


Our re-analysis of a biological data set from \cite{phylogenetics-handbook} 
using \textsc{Lev1athan} indicated that this data set is more complex
than it appears at first sight. The resulting conflicting
triplet infomation misled \textsc{Lev1athan} (and also the other two network 
construction approaches that we tried) to
postulate a too complex evolutionary scenario when using its default
option of generating a level-1 network. However \textsc{Lev1athan}'s simple
level-1 network option had no problem with this data set and
was able to correctly reconstruct the data set's widely accepted
evolutionary scenario.

\begin{figure}[H]
\centering
\includegraphics[scale=0.7]{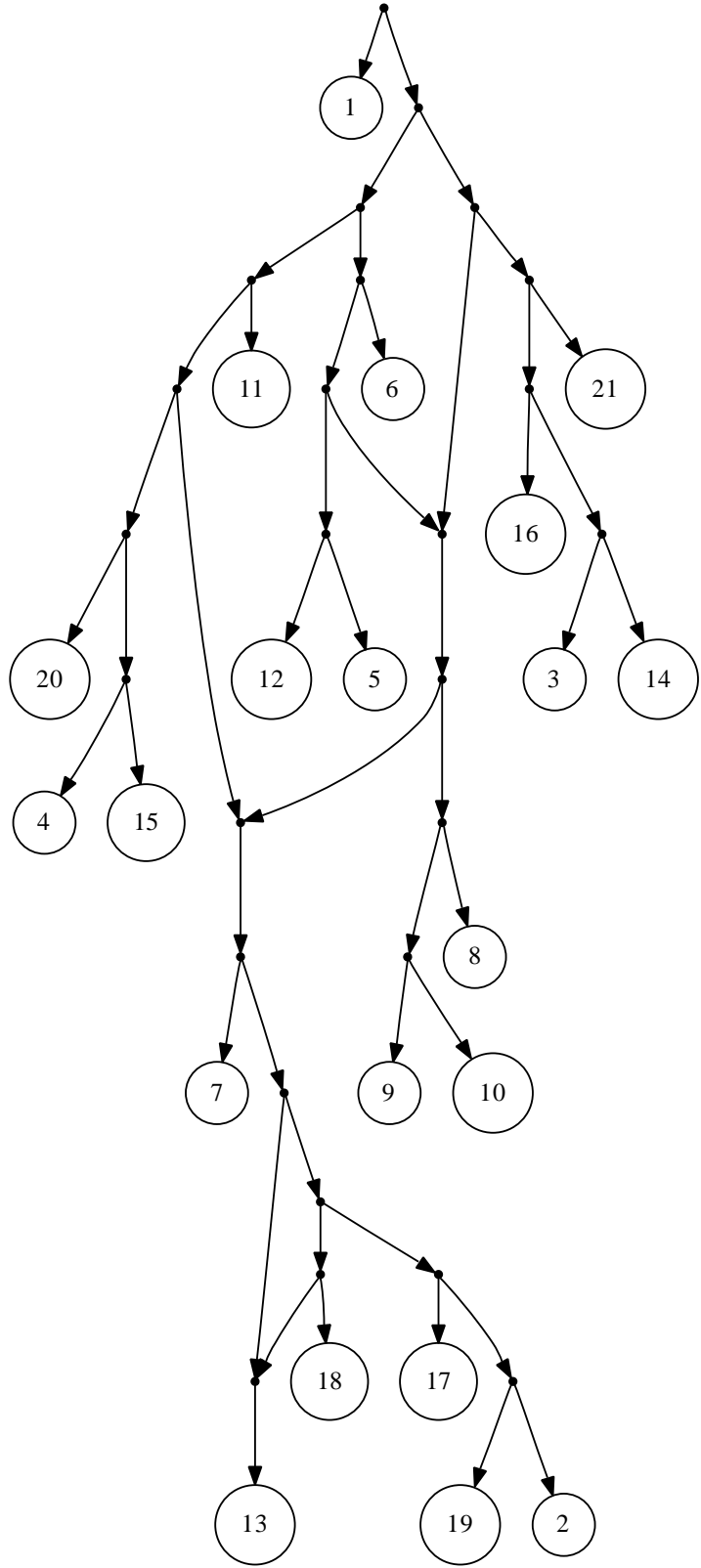}
\caption{ \label{fig:level2original}
A level-2 network on \{$1,\ldots,21$\} constructed by the \textsc{LEVEL2} algorithm \cite{lev2TCBB} for a blinded yeast dataset also
described in that paper. This network is consistent with all 1330 triplets in the dataset.
}
\end{figure}

\begin{figure}[H]
\centering
\includegraphics[scale=0.7]{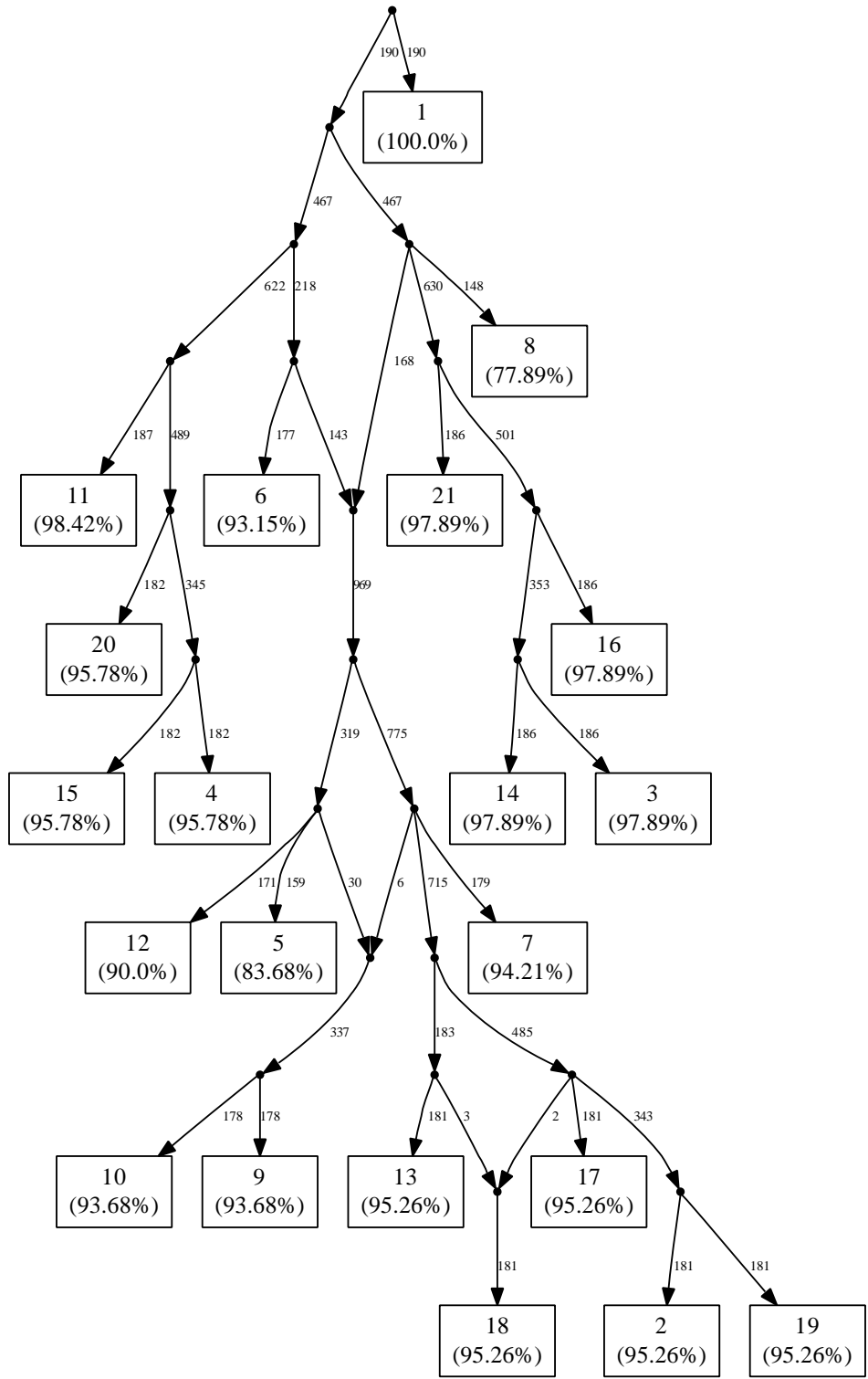}
\caption{ \label{fig:level2again}
The level-1 network constructed by \textsc{Lev1athan} when given the same dataset as discussed in Figure \ref{fig:level2original}.
This network is consistent with $94.28\%$ of the 1330 input triplets.} 
\end{figure}

To understand better how well \textsc{Lev1athan} performs, it will be necessary 
to compare it directly with an alternative method for network construction 
that uses similar input and also produces a level-1 network. Such methods are lacking at the moment.
Similarly, it is at the moment difficult to draw conclusions regarding 
the biological meaning of measures such as the $\mu$-distance, which we used in our simulation study. Comparison should
also be made between \textsc{Lev1athan} and other programs when the input is strictly 
simpler, or strictly more complex, than level-1 networks. Figures \ref{fig:level2original} and
\ref{fig:level2again} already provide interesting insights. Figure \ref{fig:level2original}
shows a level-2 network created by the \textsc{Level2} algorithm of \cite{lev2TCBB} which
is consistent with all 1330 triplets in the yeast dataset discussed in that same article.
For the same dataset \textsc{Lev1athan} constructs the level-1 network in Figure \ref{fig:level2again};
this is consistent with $94.28\%$ of the 1330 triplets. Both networks cluster the taxa together similarly;
the main difference is in the \textsc{Lev1athan} network taxa 12 and 5 have been pushed further away
from the root, whilst taxon 8 has risen closer to the root. However which one of these two networks 
is biologically more relevant is not immediately clear. In any case it should once again be noted that \textsc{Lev1athan} is in many regards much more flexible
than the \textsc{Level2} algorithm (and related algorithms such as \cite{tohabib2009, SIM08, simplicityAlgorithmica}) because \textsc{Lev1athan} always quickly 
returns a network and it does not require the input triplets to be dense or fully consistent with the output network. (In particular, the authors of \cite{lev2TCBB} had to 
use time-consuming brute-force techniques to first find a subset of the taxa that induced a triplet set fully consistent with a level-2 network).

It is also necessary to look more 
deeply at the underlying mathematics of \textsc{Lev1athan}. For 
example, the partitioning
strategy at its heart (i.e. Phase 1) is a modification 
of a strategy that was originally 
optimized for worst-case performance, not average-case performance. Yet this
strategy also seems to perform surprisingly well in the average case. 
Understanding why this is, and developing a better appreciation 
for the theoretical limits 
and strengths of triplet methods as a mechanism for 
reconstructing network topologies, remains a fascinating area of research.

\section{Acknowledgements}
We thank Vincent Moulton for many very helpful conversations during the writing of this paper. Leo van Iersel was funded by the Allan Wilson Centre for Molecular Ecology and Evolution
and Steven Kelk by a Computational Life Sciences grant of The Netherlands Organisation for Scientic Research (NWO).

\bibliographystyle{abbrv}
\bibliography{level1_arxiv}

\end{document}